\documentclass[aps,prd,reprint,twocolumn,preprintnumbers,floatfix,nofootinbib]{revtex4-1}

\usepackage[T1]{fontenc} 

\usepackage{graphicx}
\usepackage{epstopdf}
\usepackage{mathrsfs}
\usepackage{amssymb}
\usepackage{verbatim}
\usepackage{color}
\usepackage[dvipsnames]{xcolor}
\usepackage{multirow}
\usepackage{amsmath}
\usepackage{subfig}
 \usepackage{slashed} 
\usepackage{float}
\usepackage[normalem]{ulem}
\usepackage{sidecap}
\usepackage{hyperref}
\usepackage{cancel}
\usepackage{enumerate}
\hypersetup{pdfstartview=FitV,colorlinks=true,linkcolor=blue,citecolor=red,filecolor=black,urlcolor=blue}

\newcommand{\beq}{\begin{equation}}
\newcommand{\eeq}{\end{equation}}
\newcommand{\bea}{\begin{eqnarray}}
\newcommand{\eea}{\end{eqnarray}}

\newcommand{\mrm}{\mathrm}

\def\dm{\text{\tiny{DM}}}

\def\dm{\ttiny{DM}}

\def\to{\rightarrow}

\def\dm{{raritron}}

\usepackage[parfill]{parskip}
\begin{document}\sloppy 

\preprint{UMN--TH--3919/20, FTPI--MINN--20/18}
\preprint{IFT-UAM/CSIC-20-84}

\vspace*{1mm}

\title{The case for decaying spin-3/2 dark matter} 

\author{Marcos A.~G.~Garcia$^{a}$}
\email{marcosa.garcia@uam.es}
\author{Yann Mambrini$^{b}$}
\email{yann.mambrini@ijclab.in2p3.fr}
\author{Keith A. Olive$^{c}$}
\email{olive@umn.edu}
\author{Sarunas Verner$^{c}$}
\email{nedzi002@umn.edu}
\vspace{0.5cm}

 \affiliation{
$^a$
Instituto de F\'{i}sica Te\'{o}rica (IFT) UAM-CSIC, Campus de Cantoblanco, 28049 Madrid, Spain} 

\affiliation{${}^b $ Universit\'e Paris-Saclay, CNRS/IN2P3, IJCLab, 91405 Orsay, France}

\affiliation{$^c$William I. Fine Theoretical Physics Institute, School of
 Physics and Astronomy, University of Minnesota, Minneapolis, MN 55455,
 USA}

\date{\today}

\begin{abstract} 
We consider a spin-$\frac32$ particle and show that with a Planck reduced coupling, we can obtain a sufficiently long lifetime making the spin-$\frac32$
particle a good dark matter candidate.
We show that this dark matter candidate can be
produced 
during inflationary reheating through the
scattering of Standard Model particles.  The relic abundance as determined by {\em Planck} and other experimental measurements is attained for reasonable values of the reheating temperature $T_{\rm RH} \gtrsim 10^{8}$ GeV.  We consider two possible gauge invariant couplings to the extended Standard Model. 
We find a large range of masses are possible which respect the  experimental limits on its decay rate.
We expect smoking-gun signals in the form of a monochromatic photon with a possible monochromatic neutrino, which can be probed in the near future in IceCube and other indirect detection experiments.  
\end{abstract}

\maketitle

\setcounter{equation}{0}

\section{Introduction}

Almost 90 years ago, in 1933, F. Zwicky published a work that shed the 
first light on the presence of dark matter in the Coma cluster \cite{Zwicky:1933gu}, confirmed by Babcock in his Ph.D. thesis while measuring the rotation curves of Andromeda \cite{babcock}. Several studies, including the study of the stability of large-scale structures \cite{Ostriker:1973uit} confirmed the hypothesis of a dark component in the Universe. Dark matter composed of a new weakly interacting massive neutral particle (WIMP) was proposed by Steigman {\it et al.}~\cite{Gunn:1978gr} in 1978, and its precise
abundance determination was made from cosmic microwave background measurements by the {\em Planck} satellite \cite{planck} and other experiments. Despite being attractive, the WIMP paradigm is in tension with direct detection measurements (see \cite{Arcadi:2017kky} for a review). Indeed, the limit on the WIMP-nucleon scattering cross section is $\sigma_{\chi\text{-}p} \lesssim 10^{-46} \,  \rm{cm^2}$ for $m_\chi = 100$ GeV \cite{XENON,LUX,PANDAX}. The next generation of experiments will probe cross sections as low as $\sigma_{\chi\text{-}p} \lesssim 10^{-48} \, \rm{cm^2}$ \cite{Aalbers:2016jon}, approaching the irreducible neutrino background \cite{nuback}, which correspond to a Beyond the Standard Model (BSM) scale of roughly 1 PeV, i.e.~significantly above the electroweak scale. The WIMP paradigm is based on the supposition that the dark matter was initially in thermal equilibrium with the Standard Model (SM) sector before decoupling (freeze-out).
As such, its lack of dependence on initial conditions remains attractive. However, its lack of discovery to date may be implying that out-of-equilibrium processes dominate the production of dark matter. The freeze-in paradigm \cite{fimp,Bernal:2017kxu} is an interesting alternative.

The main idea behind freeze-in is that the dark sector is highly secluded from the visible sector. This seclusion may be due to a
coupling so small so as to prevent the dark matter from
equilibrating with the SM thermal bath. Unlike a WIMP and thermal freeze-out, any dark matter produced this way will have its abundance frozen in. An early example of a such a candidate is the gravitino \cite{gravitino}, produced through thermal scattering during reheating \cite{bbb,Moroi:1995fs,Ellis:2015jpg}, yet never achieving equilibrium.
There are of course many other options which include the existence of a very heavy mediator, above the maximum temperature reached during reheating, but below the Planck scale, which also prevents the dark sector from coming into thermal equilibrium with the primordial plasma.

This framework is quite common in SO(10)-like models 
\cite{Mambrini:2013iaa,MO}, high-scale supersymmetry (SUSY) \cite{Benakli:2017whb,grav2,grav3}, 
moduli portals \cite{Chowdhury:2018tzw}, spin-2 portals \cite{Bernal:2018qlk}, $Z'$ portals \cite{Bhattacharyya:2018evo} or other types of heavy mediators \cite{HighlyDecoupled}. Depending on the specific model, the production rate may be sensitive to the details of reheating, and in particular, to the effects of non-instantaneous reheating \cite{Giudice:2000ex,Ellis:2015jpg,Garcia:2017tuj,Garcia:2020eof,Bernal:2020bfj,Bernal:2020gzm}, thermalization \cite{Garcia:2018wtq}, contributions from inflaton decay \cite{Kaneta:2019zgw} or the details of the inflaton potential leading to reheating \cite{Bernal:2019mhf,Garcia:2020eof,bernal}.

Of course, the specific identity of the dark matter candidate can significantly affect its production rate. 
Nature has provided us with a spin-0 particle, spin-$\frac{1}{2}$ matter fields, spin-1 gauge fields, and a spin-2 graviton.  Is a spin-$\frac{3}{2}$ dark matter particle the missing piece in the puzzle?
Of course, the gravitino appears naturally in local supersymmetry, or supergravity, and as remarked above, was even one of the first dark matter candidates ever proposed. 
However, when Rarita and Schwinger in 1941 \cite{Rarita:1941mf} decided to simplify the (overly general) Fierz-Pauli framework \cite{Fierz:1939ix}, proposing a Lagrangian for a free spin-$\frac{3}{2}$ field, they were obviously not motivated by any arguments based on supersymmetry.

It is well known that a massive spin-$\frac{3}{2}$
particle directly coupled to a $U(1)$ gauge field
could potentially generate some acausal pathologies \cite{velo} if not treated correctly in a coherent UV
framework. This is in fact the case for any model including particles with spin $>1$. For example in ${\cal N}=2$ extended gauged supergravity, the superluminal propagation of the graviphoton is cured by gravitational backreaction. It is also possible to consider a non-minimal Rarita-Schwinger Lagrangian \cite{Porrati:2009bs}, by adding non-minimal gauge invariant terms in the action. In any case, we do not consider a spin-$\frac{3}{2}$ particle which is charged under a $U(1)$ symmetry and the potential issues raised in \cite{velo} do not apply to our work.

There have been several studies of spin-$\frac{3}{2}$ WIMP-like dark matter candidates. Effective operators coupling spin-$\frac{3}{2}$ dark matter to the Standard Model were considered for annihilations (in a freeze-out scenario and indirect detection) and scatterings (for direct detection) in \cite{Yu:2011by}. Effective interactions for spin-$\frac{3}{2}$
dark matter were also considered in \cite{eff}.
Spin-$\frac{3}{2}$ dark matter has been recently explored in \cite{Chang:2017dvm} where they proposed a WIMP-like candidate in a Higgs-portal scenario.   The detection rate in colliders was considered in  \cite{Christensen:2013aua}.
For other recent work see \cite{Savvidy:2012qa}. 
In every case, the spin-$\frac32$ dark matter is protected by a $Z_2$ symmetry to stabilize it as it is otherwise assumed to have weak scale interactions. 

In our work, we propose an extremely simple and minimal setup, where the SM model is extended with a right-handed neutrino sector needed for the seesaw mechanism for neutrino masses \cite{seesaw}.
The spin-$\frac{3}{2}$ dark matter is coupled to a single fermion and gauge field strength. The fermion is a SM singlet right-handed neutrino, and therefore, the only SM choice for the gauge field is the hypercharge gauge boson $B_\mu$. This is necessarily a dimension-5 operator and is suppressed
by a BSM scale which permits a long lifetime while at the same time serves as a portal to the SM through the left-right mixing in the neutrino sector. 
We also include an explicit lepton number violating
dimension-5 operator coupling the spin-$\frac32$ to the SM Higgs and lepton doublets as allowed by gauge invariance.
We consider each of these couplings separately. 
We show that for a large part of the parameter space, it is possible to satisfy the lifetime constraints {\it and} obtain a sufficient (and not excessive) relic density through production during reheating. 

The paper is organized as follows. We introduce the model in Sec.~II. The Lagrangian of interest will include the aforementioned dimension-5 operators and the neutrino sector giving rise to the seesaw mechanism \cite{seesaw}. In Sec.~III, we consider first the dark matter lifetime.  Assuming that the dark matter mass, $m_{3/2}$, is less than the mass of the right-handed neutrino, $M_R$, the dark matter can decay into a light neutrino and gauge boson. We then consider the production of dark matter through scattering during reheating and directly from inflaton decay.   The allowed parameter space of the model is examined in Sec.~IV, and we consider the observational signatures of the model in Sec.~V.  We summarize in Sec.~VI.

\section{The model}

\subsection{Motivations}

A massive spin-$\frac{3}{2}$ particle is described by the Rarita-Schwinger Lagrangian\footnote{Our metric convention is $g_{\mu\nu}={\rm diag}(+1,-1,-1,-1)$. In Appendix \ref{app:RS}, we provide a simple derivation of the Rarita-Schwinger Lagrangian.} \cite{Rarita:1941mf}
\beq
{\cal L}^0_{3/2}=-\frac{1}{2} \bar{\Psi}_{\mu}\left(i \gamma^{\mu\rho\nu}\partial_{\rho} + m_{3/2} \gamma^{\mu\nu}\right)\Psi_{\nu}\,,
\label{Eq:rarita}
\eeq
where $\gamma^{\mu \nu} = \gamma^{[\mu}\gamma^{\nu]}= \frac12 [\gamma^\mu,\gamma^\nu]$ and $\gamma^{\mu \nu \rho} = \gamma^{[\mu} \gamma^\nu \gamma^{\rho]}$. 
One can extract the equations of motion describing a spin-$\frac{3}{2}$ particle 
\beq
 i\gamma^{\mu\nu \rho} \partial_\nu \Psi_\rho +
 m_{3/2}\gamma^{\mu\nu} \Psi_\nu =0,~~~\gamma^{\mu \nu} \partial_{\mu} \Psi_\nu =0,
\label{cond}
\eeq
along with an extra condition appropriate for a spin-1 field
\beq
\partial^\mu \Psi_\mu =0,
\eeq
which can be deduced from the preceding constraints.
We emphasize that the condition $\gamma^\mu \Psi_\mu =0$ severely limits 
the operators available to couple a spin-$\frac{3}{2}$ field to the Standard Model sector as we discuss below.

\subsection{The Lagrangian}

In any UV completion of the SM which contains local 
supersymmetry, there is at least one spin-$\frac32$ particle, the gravitino, with well-defined couplings to SM fields.
These are governed by gauge invariance and perhaps other
symmetries, such as $R$-parity, that the full theory contains. 
In a non-supersymmetric theory, or in a theory with high-scale supersymmetry \cite{Benakli:2017whb,grav2,grav3} at mass scales below that of the superpartners,
the couplings of any spin-$\frac32$ field become greatly limited. 

Here, we do not consider a specific UV theory which contains the spin-$\frac32$ field, and therefore we use gauge invariance
to define any possible coupling to the SM which includes a right-handed and/or sterile neutrino sector. 
Our beyond the SM Lagrangian can be defined by
\beq
{\cal L}_{\rm BSM} = {\cal L}_{3/2} + {\cal L}_\nu.
\eeq
where ${\cal L}_{3/2}$ and ${\cal L}_\nu$ are defined below.

The most general Lagrangian coupling of a spin-$\frac32$
field to the SM including a right-handed and/or  sterile neutrino, $\nu_R$, consistent with Lorentz and gauge invariance is
\begin{eqnarray}
{\cal L}_{3/2} & = & i\frac{\alpha_1}{2 M_P} \bar \nu_R \gamma^\mu [\gamma^\rho,\gamma^\sigma] \Psi_\mu F_{\rho \sigma}
+ {\rm h.c.} \nonumber \\
 & & + i\frac{\alpha_2}{{2}M_P} i \sigma_2 (D^\mu H)^* \bar L  \Psi_\mu  \, .
\label{Eq:l32}
\end{eqnarray}
It is straightforward to show that dimension-4 operators of the type $\bar L H \gamma^\mu \Psi_\mu$ or dimension-5 operators $\bar \nu_R \gamma^\mu \Psi_\mu |H|^2$ are the only other Lorentz and gauge invariant operators and vanish due to the constraint given in Eq.~(\ref{cond}).
In Eq.~(\ref{Eq:l32}), $F_{\mu \nu} = \partial_\mu B_\nu - \partial_\nu B_\mu$ is 
the field strength of the Standard Model hypercharge gauge boson, $B_\mu$, $H$ is the SM Higgs doublet and $L$ a SM lepton doublet.
Due to the unknown and surely model-dependent origin of the spin-$\frac32$ field,  we have scaled its couplings in 
${\cal L}_{3/2}$
by $M_P^{-1}$ (where $M_P = 2.4 \times 10^{18}$ GeV is the reduced Planck mass), and therefore we can allow the couplings, $\alpha_i$, to take values larger as well as smaller than 1. 

Both couplings in Eq.~(\ref{Eq:l32}) are gauge and Lorentz invariant, and can be seen 
as low energy couplings for gravitino dark matter in high-scale SUSY constructions.\footnote{We will not develop this analogy any further as we prefer to remain as general as possible. We note for example, in the $\mu\nu$SSM the right-handed neutrino can mix with the Bino and generate the $\alpha_1$ coupling \cite{Choi:2009ng} and an $R$-parity violating coupling of the type $LH$ would generate the $\alpha_2$ coupling.} It is important to note that if a model contains a SM singlet such as a right-handed (or sterile) neutrino, in the absence of symmetry which prevents it, the  coupling scaled by $\alpha_1$ is present. 
In supersymmetric models, $R$-parity would prevent both couplings in Eq.~(\ref{Eq:l32}). If $R$-parity is broken, signatures of gravitino dark matter are typically a $\gamma \nu$ final state, as will be the case here.

In addition, we include a contribution to the Lagrangian
which can accommodate the seesaw mechanism~\cite{seesaw} for neutrino masses when right-handed neutrinos are included. This part of the Lagrangian is commonly written as
\beq
{\cal L}_\nu = y H \bar \nu_L \nu_R + \frac{M_R}{2} \bar \nu_R^c \nu_R + {\rm h.c.} \, ,
\label{Lnu}
\eeq
where the first term provides a Dirac mass when the SM Higgs picks up a vacuum expectation value, and the second term
provides a Majorana mass, $M_R$ for the right-handed neutrino.
 Even if $M_R \gtrsim m_{3/2}$, a coupling of the type in Eq.~(\ref{Lnu}) will generate three- and/or four-body decays of $\Psi_\mu$. 
As a consequence, spin-$\frac{3}{2}$ dark matter
is naturally unstable.
 We will refer to our metastable candidate as the \dm, an obvious tribute to the Rarita-Schwinger field.\footnote{It is interesting to note that the article immediately following the original work of  Rarita-Schwinger \cite{Rarita:1941mf}, computed the $\beta$-decay spectrum of a spin-$\frac{3}{2}$ neutrino \cite{Kusaka}. This followed Oppenheimer's suggestion \cite{opp} that the neutrino may have a spin other than $\frac{1}{2}$.
 }

The Yukawa term $y H \bar \nu_L \nu_R$ generates mixing between the neutral ($\nu_R$) and the charged ($\nu_L$) neutrino sectors, and one can define the  mass eigenstates
\bea
&&
\nu_1 = \cos \theta~ \nu_L - \sin \theta~ \nu_R
\\
&&
\nu_2= \sin \theta ~\nu_L + \cos \theta ~\nu_R,
\eea
with   
\bea
&&
m_1 = \frac{y^2 v^2}{2 M_R}; ~~~~m_2 \simeq M_R;
\nonumber
\\
&&
\tan \theta = \sqrt{\frac{m_1}{m_2}} \simeq \frac{y v}{\sqrt{2} M_R},
\label{Eq:m1}
\eea
where we have assumed $M_R \gg m_1$ (which corresponds to a classical 
seesaw mechanism of type I) and $v \simeq 246$ GeV is the vacuum expectation value of the Standard Model Higgs boson.
We have considered for simplicity only one active neutrino generation, and 
the extension to three families is straightforward. $\theta$ represents the mixing between the two sectors, and is expected to be small for large values of $M_R$,
consistent with recent limits on $m_1$ ($m_1 \lesssim 0.15$ eV~\cite{sumnu}). 

\section{The constraints}

The Lagrangian we consider contains two gauge invariant
operators with couplings $\alpha_1$ and $\alpha_2$.
In general, the contribution to the raritron decay rate
is dominated by the term proportional to $\alpha_2$
as there is no suppression from the heavy right-handed neutrino mass. However, we examine the consequences of 
each of these terms separately.
We begin with the case where $\alpha_2$ is sufficiently small that it can be neglected. In the subsequent subsection, we consider the case where the rates are dominated by $\alpha_2$ and neglect the contributions from $\alpha_1$.  

\subsection{Constraints from $\alpha_1$}

\subsubsection{The lifetime ($\alpha_1$)}

\begin{figure}[t]
\centering
\includegraphics[width=\columnwidth]{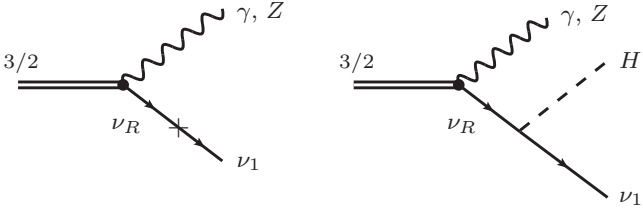}
\caption{\em \small The allowed two- and three-body decays of the \dm. }
\label{Fig:feynman1}
\end{figure}

Depending on its mass, the dominant decay channel for the \dm, $\Psi_\mu$, may contain either two or three final states. The two-body decay channel $\Psi_\mu \rightarrow \nu_1 A_\mu$ is always available. For $m_{3/2} > m_Z$, the $\nu_1 Z_\mu$ final state is open and the two-body final state dominates for $m_{3/2} < 2\pi \sqrt{15} v \simeq 6$ TeV. When $m_{3/2} \gtrsim m_H (m_H + m_Z)$, the channel $\Psi_\mu \rightarrow \nu_1 H A_{\mu} (Z_\mu)$ opens up as seen in Fig.~\ref{Fig:feynman1}. For $m_{3/2} \gtrsim 6$ TeV, the three-body final state dominates the decay width.

The decay rates for the two-body decays $\Psi_{\mu} \rightarrow A_{\mu} \nu_1$ and $\Psi_{\mu} \rightarrow Z_{\mu} \nu_1$ are 
\begin{align}
    \nonumber
    \Gamma(\Psi_{\mu} \rightarrow A_{\mu} \nu_1) & = \frac{\alpha_1^2 y^2}{8 \pi}\frac{v^2 m_{3/2}^3 \cos^2{\theta_W}}{M_P^2 M_R^2} \,, \\
    \Gamma(\Psi_{\mu} \rightarrow Z_{\mu} \nu_1) & = \frac{\alpha_1^2 y^2}{8 \pi}\frac{v^2 m_{3/2}^3 \sin^2{\theta_W}}{M_P^2 M_R^2} f \left(\frac{m_Z}{m_{3/2}} \right) \,,
\end{align}
where $f(x) = 1 - \frac{4}{3}x^2 + \frac{1}{3}x^8$ and $\theta_W$ denotes the Weinberg angle. The processes that produce the antineutrinos have the same decay rate, i.e.~$\Gamma(\Psi_{\mu} \rightarrow A_{\mu} \nu_1) = \Gamma(\bar{\Psi}_{\mu} \rightarrow A_{\mu} \bar{\nu}_1)$ and $\Gamma(\Psi_{\mu} \rightarrow Z_{\mu} \nu_1) = \Gamma(\bar{\Psi}_{\mu} \rightarrow Z_{\mu} \bar{\nu}_1 )$. 

In the limit $m_{3/2} \gg m_Z$, we find the following total two-body decay width
\begin{equation}
    \Gamma^{2b}_{3/2} \; = \; \frac{\alpha_1^2 y^2}{8 \pi}\frac{v^2 m_{3/2}^3}{M_P^2 M_R^2}.
    \label{Eq:2bodyalpha1}
\end{equation}

For the three-body decays $\Psi_{\mu} \rightarrow A_{\mu} H \nu_1$ and $\Psi_{\mu} \rightarrow Z_{\mu} H \nu_1$, we find
\begin{align} \label{eq:3ba}
    \Gamma(\Psi_{\mu} \rightarrow A_{\mu} H \nu_1) & = \frac{\alpha_1^2 y^2}{480 \pi^3} \frac{m_{3/2}^5 \cos^2{\theta_W}}{M_P^2 M_R^2} g \left(\frac{m_H}{m_{3/2}}, 0 \right)\,, \\ \label{eq:3bb}
    \Gamma(\Psi_{\mu} \rightarrow Z_{\mu} H \nu_1) & = \frac{\alpha_1^2 y^2}{480 \pi^3} \frac{m_{3/2}^5 \sin^2{\theta_W}}{M_P^2 M_R^2} g \left(\frac{m_H}{m_{3/2}}, \frac{m_Z}{m_{3/2}} \right)\, ,
\end{align}
where the expression for $g(x, y)$ is given in Appendix \ref{app:rates}. The 3-body decays to antiparticles have the same production rate $\Gamma(\Psi_{\mu} \rightarrow A_{\mu} H \nu_1) = \Gamma(\bar{\Psi}_{\mu} \rightarrow A_{\mu} H \bar{\nu}_1)$ and $\Gamma(\Psi_{\mu} \rightarrow Z_{\mu} H \nu_1) = \Gamma(\bar{\Psi}_{\mu} \rightarrow Z_{\mu} H \bar{\nu}_1)$.

In the limit $m_{3/2} \gg m_H, m_Z$, the total three-body decay rate is given by
\begin{align}
\Gamma^{3b}_{3/2} \;&=\; \frac{\alpha_1^2 y^2}{480 \pi^3} \frac{m_{3/2}^5}{M_P^2 M_R^2}\,.
\label{Eq:gamma}
\end{align}
The total two- and three-body decay rates when $m_{3/2} \gg m_H, m_Z$ correspond to lifetimes
\begin{align}
\tau^{2b}_{3/2} &\simeq 1.6 \times 10^{29} \left(\frac{10^{-2}}{y~\alpha_1} \right)^2
\left(\frac{M_R}{10^{14} \, \rm{GeV}} \right)^2  \left(\frac{10^4 \, \rm{GeV}}{m_{3/2}} \right)^3
~\mrm{s},
\nonumber
\\
\tau^{3b}_{3/2} &\simeq 5.6 \times 10^{28} \left(\frac{10^{-2}}{y~\alpha_1} \right)^2
\left(\frac{M_R}{10^{14} \, \rm{GeV}} \right)^2  \left(\frac{10^4 \, \rm{GeV}}{m_{3/2} } \right)^5
~\mrm{s.}
\label{Eq:lifetime}
\end{align}

Note that in contrast to \cite{Dudas:2014bca} (but like \cite{Dudas:2020sbq}), the four-body decay will {\it not}    dominate over the three-body decay for large values of $m_{3/2}$, because of a suppression factor of order $(m_{3/2}/M_R)^2$ between the two modes of decay.

\subsubsection{The relic abundance from scattering ($\alpha_1$)}

\begin{figure}[t]
\centering
\includegraphics[width=0.65\columnwidth]{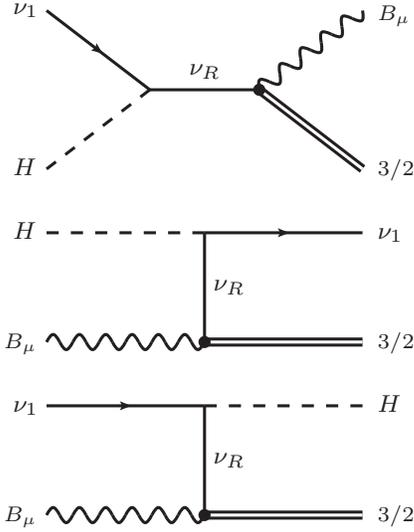}
\caption{\em \small Processes contributing to the dark matter production from the thermal bath. }
\label{Fig:feynman2}
\end{figure}

The \dm~can be produced directly from the thermal bath during reheating, which is assumed to be a result of inflaton decay.
To compute the dark matter density, $n_{3/2}$, we consider the out-of-equilibrium dark matter annihilation processes, $H + \nu_1 \rightarrow B + \Psi_{\mu}$, $H + B \rightarrow \nu_1 + \Psi_{\mu}$, and $B + \nu_1 \rightarrow H + \Psi_{\mu}$, 
as depicted in Fig.~\ref{Fig:feynman2}. We can write the Boltzmann equation as
\begin{equation}
\label{boltz1}
    \frac{d n_{3/2}}{dt} + 3 H n_{3/2} = R(T),
\end{equation}
where the Hubble parameter for the radiation-dominated Universe is given by
\begin{equation}
    H(T) = \frac{\pi \sqrt{g_*}}{\sqrt{90}} \frac{T^2}{M_P}.
\end{equation}
It is convenient to rewrite the Boltzmann equation~(\ref{boltz1}) as
\beq
\frac{d Y_{3/2}}{dT} =- \frac{R(T)}{H (T)~T^4},
\label{Eq:boltzmann}
\eeq
with $Y_{3/2} = \frac{n_{3/2}}{T^3}$.

The dark matter production rate (per unit volume per unit time) is represented by
\beq
R(T) = \frac{1}{1024 \pi^6}\int f_1 f_2 E_1 dE_1 E_2 dE_2 d\cos \theta_{12}\int |{\cal M}|^2 d\Omega_{13},
\eeq
for the processes $1+2 \rightarrow 3+4$, where $1$ and $2$ correspond to particles in the thermal bath, $3$ and $4$ correspond to produced particles, $f_1$ and $f_2$ represent the thermal distribution functions of the incoming particles, and $\cal M$ is the scattering amplitude for the processes shown in Fig.~\ref{Fig:feynman2}, with the expressions for the scattering amplitudes given in Appendix \ref{app:rates}. From these, we 
find the following dark matter production rate,
\beq
R_1(T) = \frac{ 338 \zeta(5)^2 \alpha_1^2 y^2 T^{10}}{\pi^5 M_P^2 M_R^2 m_{3/2}^2} \,,
\label{prodrate}
\eeq
where $\zeta(n)$ is the Riemann zeta function. 

It is useful to compare the raritron production rate to that of the gravitino in supersymmetric theories.
In weak scale supersymmetry, the dominant production channel is gluon + gluon $\rightarrow$ gravitino + gluino.
The dimensionful contributions to the cross section for this process
originate from the gravitino vertex $(1/m_{3/2}^2M_P^2)$, the gluino propagator ($m_{\tilde g}^2/T^4$), and $T^4$ from phase space, so that the cross section scales as $m_{\tilde g}^2/m_{3/2}^2M_P^2$. In this case, the production rate scales as $T^6 m_{\tilde g}^2/m_{3/2}^2M_P^2$.
For the case of raritron production, 
when $M_R \gg T_{\rm RH}$, the contribution from the propagator is instead $1/M_R^2$ so that the cross section scales as  $T^4/m_{3/2}^2M_P^2M_R^2$
giving a production rate which scales
as in Eq.~(\ref{prodrate}). 

Since the temperature dependence of the production rate, $T^{\beta}$, has $\beta < 12$,
the final density dark matter density is mostly sensitive to the reheat temperature, $T_{\rm RH}$, rather than the maximum temperature attained during the reheating process \cite{Garcia:2017tuj}.
Therefore, after integration of Eq.~(\ref{Eq:boltzmann}),
the density at $T_{\rm RH}$ can be written
\beq
n(T_{\rm RH})=\sqrt{\frac{2}{5g_{\rm RH}}} \frac{1014~\zeta(5)^2 \alpha_1^2 y^2}{\pi^6 M_P M_R^2 m_{3/2}^2}  T_{\rm RH}^8,
\label{ntrh}
\eeq
from which we can calculate the present relic abundance at temperature $T_0$:
\begin{align}
\Omega h^2 & \simeq 10^9 \, \frac{n(T_{\rm RH})}{\rm{cm}^{-3}}
\left(\frac{g_0}{g_{\rm RH}} \right) \left( \frac{T_0}{T_{\rm RH}}\right)^3
\left(\frac{m_{3/2}}{10^4~\mrm{GeV}} \right) \nonumber  \\
& \simeq 0.1 \left(\frac{\alpha_1}{1.1\times10^{-3}}\right)^{2}\left(\frac{427 / 4}{g_{\text {RH }}}\right)^{3 / 2}\left(\frac{T_{\text {RH }}}{10^{10} \, \mathrm{GeV}}\right)^{5}  \nonumber \\
& \quad \times \left(\frac{m_{1}}{0.15 \, \mathrm{eV}}\right)\left(\frac{10^{14} \, \mathrm{GeV}}{M_{R}}\right)\left(\frac{10^{4} \, \mathrm{GeV}}{m_{3 / 2}}\right),
\label{Eq:omegabis}
\end{align}
where $g_0 = 43/11$. 
In writing Eq.~(\ref{Eq:omegabis}), we have 
substituted Eq.~(\ref{Eq:m1}) for $y$, assuming a characteristic mass of 0.15 eV for the light neutrino. Note that Eqs.~(\ref{ntrh}) and (\ref{Eq:omegabis}) 
are derived using an instantaneous reheating approximation. 
Dropping this approximation results in a density which is about 2 times larger
for a production rate proportional to $T^{10}$ as in Eq.~(\ref{prodrate}) \cite{Garcia:2017tuj}. 

It is interesting to note that the same set of parameters which provide a sufficiently long-lived raritron so as to respect the indirect detection constraints, Eq.~(\ref{Eq:lifetime}), also produce a relic density in agreement with {\em Planck} data, Eq.~(\ref{Eq:omegabis}) for a reasonable reheating temperature $T_{\rm RH} \simeq 10^{9}$ GeV. 

The scattering processes considered in Fig.~\ref{Fig:feynman2} led to a scattering cross section that scales with the fourth power of the energy of the scatterers, $\sigma\sim s^2$ [c.f.~Eq.~(\ref{eq:MHnupsiB}) in Appendix~\ref{app:rates}]. In the classification of~\cite{Garcia:2017tuj}, this corresponds to the $n=4$ scenario ($\sigma\sim s^{n/2}$). For such a steep dependence on the energy of the scatterers, the instantaneous thermalization approximation can severely underestimate the magnitude of the relic abundance. Indeed, in~\cite{Garcia:2018wtq} it was found that the production of particles from scatterings in the not-yet-thermalized relativistic plasma, present at the earliest stages of reheating, will generically determine the dark matter abundance if $n>2$. The decay products have initial momenta $p\sim m_{\Phi}$, where $m_\Phi$ is the mass of the inflaton, and it is only after interactions in the plasma can equilibrate that $p\sim T$. The very energetic particles produced before the thermalization of the Universe can therefore dominate the dark matter density budget despite their dilution by entropy production during the late stages of reheating.

Let us assume for definiteness that the inflaton decays predominantly to Higgs bosons, and subdominantly to neutrinos and gauge bosons. When this is the case, the pre-thermal production rate of raritrons can be easily estimated, following the procedure outlined in~\cite{Garcia:2018wtq}. When reheating ends, the number density of pre-thermally generated raritrons via the processes depicted in Fig.~\ref{Fig:feynman2} can be written as
\begin{align} 
n(T_{\rm RH}) \;\simeq\; &\left(\frac{5\pi^2 g_{\rm RH}}{72}\right)^{17/10} \frac{2\alpha_1^2 y^2 m_{\Phi}^{14/5} T_{\rm RH}^{34/5} \mathcal{B}_1}{147\pi \alpha_{\rm SM}^{16/5} m_{3/2}^2 M_R^2 M_P^{13/5}}\,,
\end{align}
where
\beq
\mathcal{B}_1 \;\equiv\; {\rm Br}_{\nu_1} + \frac{2}{3} {\rm Br}_B + \frac{1}{6} {\rm Br}_{\nu_1} {\rm Br}_B\,.
\eeq
Here ${\rm Br}_{\nu_1}$ (${\rm Br}_B$) denotes the branching ratio to light neutrinos (to $B$), and $\alpha_{\rm SM}$ denotes the gauge coupling strength of the interaction responsible for thermalization during reheating. This results in the following closure fraction,
\begin{align} \notag
\Omega_{3/2}h^2 \;\simeq\; &0.1 \left(\frac{\alpha_1}{1.1\times10^{-3}}\right)^2 \left(\frac{0.030}{\alpha_{\rm SM}}\right)^{16/5} \left(\frac{m_1}{0.15\,{\rm eV}}\right)\\ \notag
& \times \left(\frac{g_{\rm RH}}{427/4}\right)^{7/10}  \left(\frac{10^4\,{\rm GeV}}{m_{3/2}}\right) \left(\frac{10^{14}\,{\rm GeV}}{M_R}\right)\\ \notag
& \times \left(\frac{m_{\Phi}}{3\times 10^{13}\,{\rm GeV}}\right)^{14/5} \left(\frac{T_{\rm RH}}{10^{10}\,{\rm GeV}}\right)^{19/5}\\
& \times \left(\frac{\mathcal{B}_1}{7\times 10^{-4}}\right)\,.
\label{nontherm}
\end{align}
Note that for the chosen model parameters this non-thermally produced population of raritrons dominates over the thermally produced one (\ref{Eq:omegabis}) if $\mathcal{B}_1 \gtrsim 7\times 10^{-4}$. This ``enhancement'' of the production rate is dependent on the possibility of producing the parent scatterers $H$, $\nu_L$ and/or $B$ directly from inflaton decay. Substantially suppressing two of these decay channels will lead to a raritron population overwhelmingly dominated by late-time reheating thermal effects, with rate (\ref{prodrate}). In the following section we specialize to reheating driven by the coupling between the inflaton $\Phi$ and $\nu_R$. In the case when $M_R\gg m_{\Phi}$, the dominant decay channel of $\Phi$ is precisely to Higgs bosons, while the decay to neutrinos is suppressed by
\beq
{\rm Br}_{\nu_1} \;\simeq\; \left(\frac{m_1 m_{\Phi}}{8 M_R^2}\right)^2 \ln^{-2}\left(\frac{M_R^2}{m_{\Phi}^2}\right)\,,
\eeq
which is $\mathcal{O}(10^{-51})$ for the fiducial values considered in (\ref{nontherm}). Moreover, we assume no direct production of gauge bosons.\footnote{The timescale for efficient emission of energetic ($p\sim m_{\Phi}$) gauge bosons from the inflaton decay products is typically larger than the thermalization timescale~\cite{Harigaya:2014waa,Harigaya:2013vwa,Mukaida:2015ria}.} This then renders non-thermal production completely negligible in this case. In what follows we will therefore disregard this production mechanism, albeit having in mind that for a different reheating process it could be of importance.

\subsubsection{The relic abundance from inflaton decay ($\alpha_1$)}

In principle, it is also necessary to consider dark matter production directly from inflaton decay. 
We parametrize the total width for inflaton decay as follows,
\beq
\Gamma_\Phi^{\rm tot} = \frac{y_\Phi^2}{8 \pi} m_\Phi\,.
\label{infdec}
\eeq
Inflaton decay produces a thermal bath, and we define the moment of reheating to be the time of inflaton-radiation equality.\footnote{We are further assuming a matter-dominated Universe prior to decay and $H = \frac{2}{3t}$.} 
During the process of reheating, the temperature of the newly created radiation bath falls as $T \propto a^{-3/8}$, where $a$ is the cosmological scale factor. From the solution to the set of Boltzmann/Friedmann equations
\bea
&&
\dot \rho_\Phi 
+ 3 
H \rho_\Phi = - \Gamma_\Phi \rho_\Phi,
\label{Eq:eqrhophi} \\
&&
\dot \rho_R + 4 H \rho_R \;=\; \Gamma_\Phi \rho_\Phi\,,
\label{Eq:eqrhor}
\\
&&
H^2 \;=\; \frac{\rho_\Phi + \rho_R}{3 M_P^2} \;\simeq\; \frac{\rho_\Phi}{3 M_P^2}\, ,
\label{Eq:eqh}
\eea
we find \cite{Garcia:2020eof}
\beq
\frac{\pi^2 g_{\rm RH} T_{\rm RH}^4}{30} \;=\; \frac{12}{25}\left(\Gamma_{\Phi}^{\rm tot} M_P\right)^2\, ,
\eeq
and we can write
\beq
T_{\rm RH} \simeq 6 \times 10^{14}~{\rm GeV} y_\Phi \left( \frac{m_\Phi}{3 \times 10^{13}~{\rm GeV}} \right) \, .
\label{Try}
\eeq
The source of the Yukawa coupling $y_\Phi$ is of course model dependent. If the inflaton is directly coupled to the SM, there may be, for example, a direct coupling of the inflaton to the Higgs of the type $\Phi HH^*$, or the decay to Standard Model fields may involve
loops containing SM and/or BSM fields. As a minimal assumption, we assume first that the 
inflaton couples directly only to the BSM field $\nu_R$ through $y_{\nu}\Phi\bar{\nu}_R \nu_R$ and this is the main source of the reheating.

\begin{figure}[t]
\centering
\includegraphics[width=4.5cm]{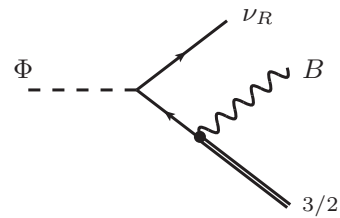} 
\caption{Three-body decay of the inflaton producing a raritron when $M_R \ll m_\Phi$.}
\label{br1}
\end{figure}

If $m_{\Phi}>M_R$, then the decay rate of $\Phi$ is simply $\Gamma_{\Phi}=y_{\nu}^2 m_{\Phi}/8\pi$ ($y_\Phi = y_\nu$), and the raritron is produced through the decay process shown in Fig.~\ref{br1}.
The partial width in this case is
\begin{equation}
    \Gamma_{\Phi \rightarrow 3/2} \;\simeq\; \frac{\alpha_1^2 y_{\nu}^2 m_{\Phi}^5}{288 \pi^3 m_{3/2}^2 M_P^2}\,,
    \label{Phidec32}
\end{equation}
where we have assumed that $m_{\Phi} \gg M_R, \, m_{3/2}$. The branching ratio is therefore given by
\beq
{\rm Br}_{3/2} = \frac{\alpha_1^2 m_\Phi^4}{36 \pi^2 m_{3/2}^2 M_P^2}\, .
\eeq

For a given branching ratio, the number density of raritrons at the end of reheating will be given by~\cite{Kaneta:2019zgw,Garcia:2020eof}
\beq
n(T_{\rm RH}) \;=\; \frac{\pi^2 {\rm Br}_{3/2} g_{\rm RH} T_{\rm RH}^4}{18 m_{\Phi}}\,,
\eeq
and the relic density in turn takes the form
\begin{align}
\Omega_{3/2} h^{2}  \;&\simeq\;  0.1 \times \left(\frac{ {\rm Br}_{3/2}}{9 \times 10^{-11}} \right) \left( \frac{T_{\rm RH}}{10^{10}~{\rm GeV}} \right) \nonumber 
\\ 
 &\quad \times  \left(\frac{3 \times 10^{13}~{\rm GeV}}{m_\Phi} \right) \left( \frac{m_{3/2}}{10^4~{\rm GeV}} \right) \label{eq:omegabrgen} \\
 &\simeq\; 0.1 \left( \frac{\alpha_1}{5 \times 10^{-9}} \right)^2 \left( \frac{T_{\rm RH}}{10^{10}~{\rm GeV}} \right)  \nonumber \\
 &\quad \times  \left(\frac{m_\Phi}{3 \times 10^{13}~{\rm GeV}} \right)^3 \left( \frac{10^4~{\rm GeV}}{m_{3/2}} \right)
 \, .
 \label{Eq:omegabrbis}
\end{align}
As one can see, a
very small coupling between the raritron and $\nu_R$ is required
to avoid overclosure.\footnote{Since the decay of the inflaton is not instantaneous, entropy production continues for some time beyond inflaton-radiation equality. A numerical calculation shows that this injection of entropy, overlooked in our analytical estimates, reduces the value of $\Omega_{3/2}$ by a factor of $\sim 0.7$.}

When $M_R > m_{\Phi}$, the direct decay to $\nu_R$ is not kinematically allowed. There is a two-body decay $\Phi \to \nu_1 \bar{\nu}_1$ and the decay rate for this channel would be given by
$y_\nu^2 \theta^4 m_\Phi/8 \pi$. However, 
decay to Higgs and light neutrino pairs can proceed through the loop diagrams shown in Fig.~\ref{brl1} and is computed in Appendix \ref{app:loops}.\footnote{There are also four-body decays with an off-shell $\nu_R$, but those rates are highly suppressed, $\Gamma_{4b} \propto y_\nu^2 y^4 m_\Phi^5/M_R^4$.} For $M_R \gg m_{\Phi}, m_{3/2}$, we find the following partial widths to Higgses and neutrinos,
\beq
\Gamma_{\Phi \rightarrow H} \;\simeq\; \frac{y_{\nu}^2 y^4 M_R^2}{256 \pi^5 m_{\Phi}}\ln^2\left(\frac{M_R^2}{m_{\Phi}^2}\right)\,,
\label{phitoH}
\eeq
and 
\begin{equation}
    \label{phitonuL}
    \Gamma_{\Phi \rightarrow \nu_1} = \frac{y_{\nu}^2 y^4 v^4 m_{\phi}}{32 \pi M_R^4} \left( 1 + \frac{y^4}{256 \pi^4} \right) \,,
\end{equation}
where in the decay to $\nu_1 \bar{\nu}_1$, we include the tree-level and 
one-loop contributions. 
In this case, the decay to Higgs is clearly dominant and we can associate (\ref{phitoH}) with a total rate such that 
$y_\Phi = (y_\nu y^2/4\pi^2) (M_R/m_\Phi) \ln (M_R/m_\Phi)^2$.

\begin{figure}[t]
\centering
\includegraphics[width=4.5cm]{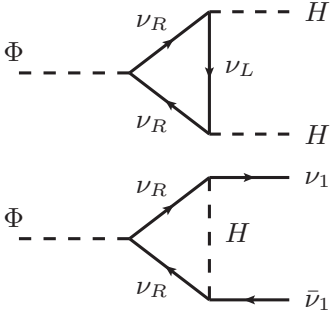}
\caption{Inflaton decay to Higgses and light neutrinos when $M_R \gg m_\Phi$.}
\label{brl1}
\end{figure}

\begin{figure}[t]
\centering
\includegraphics[width=4.9cm]{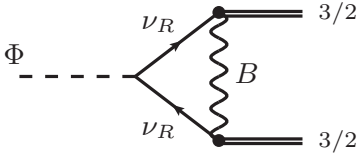} 
\caption{Inflaton decay to two raritrons when $M_R \gg m_\Phi$.}
\label{brl2}
\end{figure}
Inflaton decays to raritrons is also possible when $M_R > m_\Phi$. A tree-level decay to $\nu_1 B \Psi_{\mu}$ has a rate given by Eq.~(\ref{Phidec32}) multiplied by $\theta^2$.  There is also the loop process shown in Fig.~\ref{brl2} \footnote{We note that if there is no direct coupling between the inflaton and $\nu_R$ (i.e.~$y_\nu = 0$), and the inflaton decays directly to SM particles, such as $\Phi \to H H^*$, raritron production through inflaton decay is still possible at two loops.} and its partial width is given by
\footnote{We find an analogous suppression for the four-body decay to B and 3/2.}
\begin{equation}
 \label{phitorar}
   \Gamma_{\Phi \rightarrow 3/2} \simeq  \frac{\alpha_1^4 y_{\nu}^2 M_R^4 m_{\Phi}^5}{4 \pi^5 M_P^4 m_{3/2}^4 } \Upsilon \left(\frac{M_R^2}{m_{\phi}^2} \right) \,,
\end{equation} 
where $\Upsilon (M_R^2/m_{\phi}^2 ) = (\ln(M_R^2/m_{\phi}^2) - 5/6 )^2$.  
The loop decay dominates whenever
\beq
\alpha_1^2 M_R^4 \Upsilon \left(\frac{M_R^2}{m_{\phi}^2} \right) > \frac{\pi^2}{72} M_P^2 m_{3/2}^2.
\label{dom}
\eeq
When $M_R > m_\Phi$, the right-hand side of Eq.~(\ref{dom}) should be multiplied by $\theta^2$. When the loop dominates, 
the branching ratio is given by 
\beq
{\rm Br}_{3/2} \;\simeq\; \frac{64 \alpha_1^4 M_R^2 m_{\Phi}^6}{y^4 m_{3/2}^4 M_P^4} \frac{\Upsilon (M_R^2/m_{\phi}^2)}{\ln^2(M_R^2/m_{\phi}^2 )}
\,.
\eeq
Using Eq.~(\ref{eq:omegabrgen}) we can immediately deduce the relic abundance,
\begin{align}\notag
& \Omega_{3/2}h^2 \; \simeq\; \frac{4 \pi ^4 g_0  \alpha_1 ^4 n_{\gamma} v^4 m_{\Phi}^5  T_{\rm RH}}{9 \zeta(3) \rho_{c}h^{-2} m_1^2 m_{3/2}^3 M_P^4 } \frac{\Upsilon (M_R^2/m_{\phi}^2)}{\ln^2(M_R^2/m_{\phi}^2 )} \\ \notag
&\simeq\; 0.1 \left(\frac{\alpha_1}{1.1\times 10^{-8}}\right)^4 \left(\frac{m_{\Phi}}{3\times 10^{13}\,{\rm GeV}}\right)^5 \left(\frac{0.15\,{\rm eV}}{m_1}\right)^2\\ 
& \times  \left(\frac{10^{4}\,{\rm GeV}}{m_{3/2}}\right)^3 \left(\frac{T_{\rm RH}}{10^{10}\,{\rm GeV}}\right)
\times \frac{\Upsilon (M_R^2/m_{\phi}^2)}{\ln^2(M_R^2/m_{\phi}^2 )}\,.
\label{Eq:omegabr}
\end{align}
In this case too, a small coupling between the raritron and
$\nu_R$ is required to avoid overclosure, though for $M_R > 
m_\Phi$, it is more easily mitigated by taking a large raritron mass as $\Omega_{3/2} h^2 \propto \alpha_1^4/m_{3/2}^3$.
As one can see, for a given set of parameters ($\alpha_1$, $M_R$, $y$), the possibility of the direct production of raritrons from inflaton decay opens up a new window, allowing for the production of super-heavy spin-3/2 dark matter. Indeed the raritron mass may be well above the reheating temperature, and then {\it only} accessible through decay rather than from scattering.

Larger values of $\alpha_1$ are possible if there are inflaton decay
channels directly to the SM. Thus if $y_\Phi$ (defined in Eq.~(\ref{infdec})) is much larger than $y_\nu$. In this case,
\beq
{\rm Br_{3/2}} = \frac{2 \alpha_1^4}{\pi^4} 
\left(\frac{y_\nu}{y_\Phi} \right)^2
\frac{M_R^4 m_\Phi^4}{M_P^4 m_{3/2}^4} \Upsilon \left(\frac{M_R^2}{m_{\phi}^2} \right),
\eeq
which gives
\begin{align}
& \Omega_{3/2}h^2 \;\simeq\; \frac{g_0  \alpha_1 ^4 y_\nu^2 n_{\gamma} M_R^4 m_{\Phi}^3  T_{\rm RH}}{18 \zeta(3) \rho_{c}h^{-2} y_\Phi^2 m_{3/2}^3 M_P^4 } \Upsilon \left(\frac{M_R^2}{m_{\phi}^2} \right) \\ \notag
&\simeq\; \left( \frac{9}{40\pi^4g_{\rm RH}} \right)^{1/2} \frac{g_0  \alpha_1 ^4 y_\nu^2 n_{\gamma} M_R^4 m_{\Phi}^4}{18 \zeta(3) T_{\rm RH} \rho_{c}h^{-2} m_{3/2}^3 M_P^3 }\Upsilon \left(\frac{M_R^2}{m_{\phi}^2} \right) \\ 
&\simeq\; 0.1 \left(\frac{\alpha_1 \sqrt{y_{\nu}}}{2.7 \times 10^{-10}}\right)^4 \left(\frac{427/4}{g_{\rm RH}}\right)^{1/2} \left(\frac{m_{\Phi}}{3\times 10^{13}\,{\rm GeV}}\right)^4 \nonumber \\ \notag
& \times  \left(\frac{10^{14}\,{\rm GeV}}{M_R}\right)^4 \left(\frac{10^{4}\,{\rm GeV}}{m_{3/2}}\right)^3 \left(\frac{10^{10}\,{\rm GeV}}{T_{\rm RH}}\right) \Upsilon \left(\frac{M_R^2}{m_{\phi}^2} \right)\,,
\end{align}
where we have used (\ref{Try}) to substitute $T_{\rm RH}$ for $y_\Phi$. Even in this case, we require the product of couplings
$\alpha_1 \sqrt{y_\nu} \approx 10^{-10}$ to obtain the correct relic density.

\subsection{Constraints from $\alpha_2$}

Having established the decay and production rates for the raritron stemming from the coupling to the right-handed neutrino sector, we now repeat the analysis when
the contributions proportional to $\alpha_2$ are dominant. 
\subsubsection{The lifetime ($\alpha_2$)}

\begin{figure}[t]
\centering
    \includegraphics[width=\columnwidth]{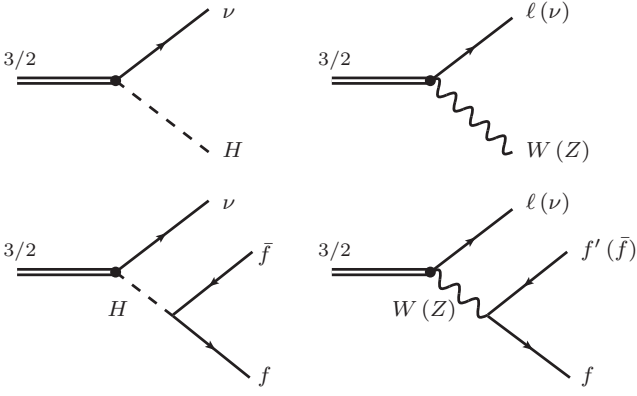}
\caption{\em \small The allowed two- and three-body decays of the \dm~ with coupling $\alpha_2$.}
\label{Fig:feynmana2}
\end{figure}

When the term proportional to $\alpha_2$ dominates raritron decay, two-body decays to  $\ell W$, $\nu Z$, and $\nu H$ are possible provided that $m_{3/2} > m_W$.\footnote{For these decay rates, unless otherwise specified, $\ell$ ($\nu$) stands for all of $e,\mu,\tau$ ($\nu_e,\nu_\mu,\nu_\tau$).} 
The Feynman graphs for these are shown in Fig.~\ref{Fig:feynmana2}.
When $m_{3/2} < m_W$ , there are many three-body final
states where $W, Z$, and $H$ are all produced off shell. 
These are also shown in Fig.~\ref{Fig:feynmana2}

The decay rates for the two-body decays $\Psi_{\mu} \rightarrow \ell (\nu)\, W (Z) $ and $\Psi_{\mu} \rightarrow  \nu H$ are 
\begin{align}
    \Gamma(\Psi_{\mu} \rightarrow Z_{\mu} \nu) & = \frac{\alpha_2^2}{3072 \pi}\frac{m_{3/2}^3}{M_P^2} 
    f_1 \left(\frac{m_Z}{m_{3/2}} \right) \,, \\
    \Gamma(\Psi_{\mu} \rightarrow H \nu) & = \frac{\alpha_2^2}{3072 \pi}\frac{m_{3/2}^3}{M_P^2} f_2 \left(\frac{m_H}{m_{3/2}} \right) \,, 
\end{align}
where $f_1(x) = (1-x^2)^2(1 + 10 x^2 + x^4)$ and $f_2(x) = (1-x^2)^4$. The rate for $\Psi_{\mu} \rightarrow \ell W $ is twice that of the $\nu_1 Z$ final state.

In the limit $m_{3/2} \gg m_H$, we find the following total two-body decay width
\begin{equation}
    \Gamma^{2b}_{3/2} \; = \; \frac{\alpha_2^2}{256 \pi}\frac{m_{3/2}^3}{M_P^2}\, .
    \label{Eq:2bodyalpha2}
\end{equation}
Comparing Eqs.(\ref{Eq:2bodyalpha1}) and (\ref{Eq:2bodyalpha2}) we notice that for $\alpha_2> 4\sqrt{2} \frac{y v}{M_R} \alpha_1$, this contribution to the two-body decays dominates the raritron lifetime.

For the three-body decay to fermions via Higgs exchange ($m_H \gg m_{3/2}$), we find
\begin{equation}
    \Gamma(\Psi_{\mu} \rightarrow \nu \bar{f} f) \simeq \frac{\alpha_2^{2} N_{cf} h_{f}^{2} m_{3 / 2}^{7}}{1474560 \pi^{3} m_{H}^{4} M_{P}^{2}} g_1\left(\frac{m_f}{m_{3/2}} \right),
\end{equation}
where the Yukawa coupling is $h_f = m_f \sqrt{2}/v$, $N_{cf}$ is the number of colors of fermion $f$, and the expression for $g_1(x)$ is given in Appendix~\ref{app:rates}.
If $m_{3/2} \gg m_H \gg m_f$, we have
\begin{equation}
    \Gamma(\Psi_{\mu} \rightarrow \nu \bar{f} f) \simeq \frac{\alpha_2^{2} N_{cf}  h_{f}^{2} m_{3 / 2}^{3} m_{H}}{49152 \pi^{2} M_{P}^{2} \Gamma_{H}} \, ,
\end{equation}
which is equal to the product $\Gamma_{\Psi_{\mu}\rightarrow H\nu}\times {\rm Br}_{H\rightarrow \bar{f}f}$.

For the three-body decay to fermions via $W$-boson exchange (assuming $m_{W} \gg m_{3 /2} \gg m_{f}, m_{f'}, m_{\ell}$), we find
\begin{equation}
    \Gamma \left({\Psi_{\mu} \rightarrow \ell f f^{\prime}} \right) \simeq \frac{\alpha_2^{2} g^2 N_{cf}\left|V_{f f^{\prime}}\right|^{2} m_{3 / 2}^{5}}{61440 \pi^{3} m_W^{2} M_{P}^{2}},
\end{equation}
where $V_{f f^{\prime}}=\delta_{f f^{\prime}}$ for leptonic decays, and $V_{f f^{\prime}}$ is the Cabibbo-Kobayashi-Maskawa matrix element for decays into quarks.
If $m_{3/2} \gg m_W \gg m_{f, f'}$, we have
\begin{equation}
    \Gamma \left({\Psi_{\mu} \rightarrow \ell f f^{\prime}} \right) \simeq \frac{\alpha_2^{2} g^2 N_{c}\left|V_{f f^{\prime}}\right|^{2} m_{3 / 2}^{3} m_{W}}{73728 \pi^{2}  M_{P}^{2} \Gamma_{W}}\, ,
\end{equation}
equal to $\Gamma_{\Psi_{\mu}\rightarrow W\ell}\times {\rm Br}_{W\rightarrow f'f}$.

For the three-body decay to fermions via $Z$~exchange ($m_Z \gg m_{3/2}$), we find
\begin{equation}
    \Gamma(\Psi_{\mu} \rightarrow \nu \bar{f} f) \simeq \frac{\alpha_2^{2} (g^2+{g^{\prime}}^2) N_{cf} m_{3 / 2}^{5}}{122880 \pi^{3} m_Z^{2} M_{P}^{2}} g_2 \left(\frac{m_{f}}{m_{3/2}}\right),
\end{equation}
where $g_2$ is given in Appendix~\ref{app:rates}.
If $m_{3/2} \gg m_Z \gg m_f$, we find
\begin{equation}
    \Gamma(\Psi_{\mu} \rightarrow \nu \bar{f} f) \simeq \frac{\alpha_2^{2} N_{cf} \left( g^2 + {g^\prime}^2 \right) \left(c_{A}^{2}+c_{V}^{2}\right) m_{3 / 2}^{3} m_{Z}} {147456 \pi^{2} M_{P}^{2} \Gamma_{Z}}\,,
\end{equation}
where $c_V=T_3-2Q\sin^2\theta_W$, $c_A=T_3$ denote the axial and vector couplings of $f$. This rate is equal to $\Gamma_{\Psi_{\mu}\rightarrow Z\nu}\times {\rm Br}_{Z\rightarrow \bar{f}f}$. 

\begin{figure}[t]
\centering
\includegraphics[width=\columnwidth]{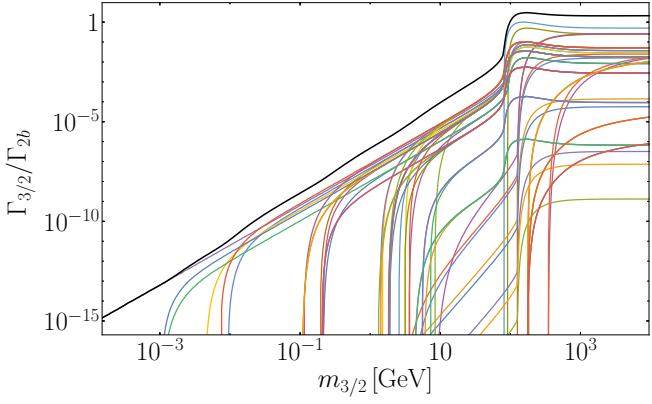}
\caption{\em \small The allowed partial rates for both two-body and three-body final states relative to the total two-body decay rate as a function of the raritron mass. The black line corresponds to the sum of all partial rates.  }
\label{fig:decays}
\end{figure}

When $m_{3/2} \ll m_e$, we can approximate the total decay rate as
\begin{equation}
    \Gamma^{3 b}_{3/2}=9 \Gamma_{\Psi_{\mu} \rightarrow \nu_{\ell} \bar{\nu}_{\ell^{\prime}} \nu_{\ell^{\prime}}}=\frac{3 \alpha_2^{2}(g^2+{g^{\prime}}^2) m_{3 / 2}^{5}}{81920 \pi^{3} m_Z^{2} M_{P}^{2}}.
    \label{3brate1}
\end{equation}
The various two- and three-body partial rates are shown relative to the total two-body rate in Fig.~\ref{fig:decays}.

When $m_{3/2} < m_W$, the raritron lifetime is solely determined by the three-body rates, and the number of 
different fermion final state pairs will depend on the raritron mass. This is clearly seen in Fig.~\ref{fig:decays} as more partial rates come into play as $m_{3/2}$ is increased.  
Note that for $m_{3/2} \ll m_W$, (and thus $\ll m_t$) the three-body decay through Higgs exchange is subdominant. Indeed, we find
\beq
\frac{\Gamma^{3b/H}_{3/2}}{\Gamma^{3b/W}_{3/2}} \propto \frac{h_f^2 m_{3/2}^2 M_W^2}{g^2 M_H^4} \ll 1 \, .
\eeq

The total rate can be expressed by approximate analytic expression in three separate mass regimes. For $m_{3/2} > m_H$, the sum of all
the three-body final state channels gives a rate slightly larger
(by a factor of 1.09) than the two-body rate so that the lifetime for large masses is 
\begin{align}
\frac{\tau_{3/2}}{10^{28} {\mrm s}} &\simeq 14.8  \left(\frac{10^{-7}}{\alpha_2} \right)^2
 \left(\frac{1 \, \rm{GeV}}{m_{3/2}} \right)^3; \qquad m_{3/2} > m_H.
\label{Eq:lifetime22b}
\end{align}
For $m_{3/2} < m_W$, the two-body rate is kinematically forbidden,
and the total three-body rate is 
complicated by the various thresholds and a single simple expression for the lifetime is not possible. However, for $m_{3/2} < m_e$, we can use Eq.~(\ref{3brate1})
\begin{align}
\frac{\tau_{3/2}}{10^{28} {\mrm s}} &\simeq 4.8  \left(\frac{10^{-3}}{\alpha_2} \right)^2
\left(\frac{1 \, \rm{GeV}}{m_{3/2} } \right)^5 ; \qquad m_{3/2} < m_e
\label{Eq:lifetime23b1} ,
\end{align}
while for $m_e < m_{3/2} < m_W$
we can fit the total three-body rate given by the 
black line in Fig.~\ref{fig:decays} and find
\begin{align}
\frac{\tau_{3/2}}{10^{28} {\mrm s}} &\simeq 0.6 \left(\frac{10^{-3}}{\alpha_2} \right)^2
\left(\frac{1 \, \rm{GeV}}{m_{3/2} } \right)^{5.28} ; m_e < m_{3/2} < m_W .
\label{Eq:lifetime23b2}
\end{align}

As one can see, for the same raritron mass,
a sufficiently long lifetime requires $\alpha_2 \ll \alpha_1$ as expected. Conversely, for similar values of the couplings, the rate proportional to $\alpha_2$ 
will produce a long lifetime at small values of the rariton masses. Typically for large enough $\alpha_2$,
we require $m_{3/2} < m_W$ and only the three-body
decay channels are open. 

\subsubsection{The relic abundance from scattering ($\alpha_2$)}

\begin{figure}[t]
\centering
\includegraphics[width=\columnwidth]{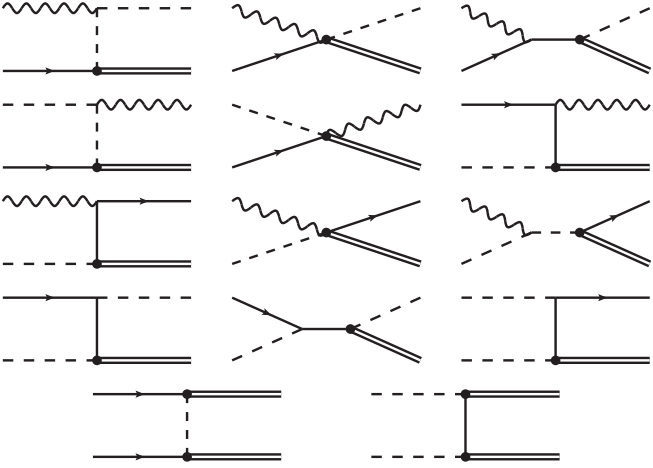}
\caption{\em \small Processes contributing to the dark matter production from the thermal bath, for the coupling $\alpha_2$. In each diagram the dashed lines denote the Higgs $SU(2)$ doublet, the wavy curves the gauge bosons $W_{\mu}$ or $B_{\mu}$, and the solid lines the $SU(2)$ doublets $L_{e,\mu, \tau}$ or the singlets $\ell_R$, as appropriate. }
\label{Fig:feynmana22}
\end{figure}

When $\alpha_2$ is the dominant coupling, there are many
channels from which the \dm~can be produced directly from the thermal bath during reheating.
These include $L + L \rightarrow \Psi_{\mu} + \Psi_{\mu}$, $H + H \rightarrow \Psi_{\mu} + \Psi_{\mu}$, $H + L \rightarrow B_\mu (W_\mu) + \Psi_{\mu}$, $H + B_\mu (W_\mu) \rightarrow L + \Psi_{\mu}$, $L + B_\mu (W_\mu) \rightarrow H + \Psi_{\mu}$, as well as many other diagrams mediated by Higgs exchange
as depicted in Fig.~\ref{Fig:feynmana22}.

The temperature dependence of the production rate 
rate will depend on the number of raritons in the final state. The total rate is
\begin{eqnarray}
R_2(T) & = & \frac{3139 \alpha_2^4 \pi^7 T^{12}}{76204800M_P^4 m_{3/2}^4 } \nonumber \\
& & + \frac{\alpha_{2}^{2} \pi^{3} \left(1521 g^2 +165 g'^2 + 441h_t^2 + 50h_{\tau}^2 \right)T^{8}}{3110400 M_{P}^{2} m_{3 / 2}^{2}} \nonumber \\
& = & \frac{0.125 \alpha_2^4  T^{12}}{M_P^4 m_{3/2}^4} + \frac{ 0.011 \alpha_2^2  T^{8}}{ M_P^2 m_{3/2}^2} \,,
\label{prodrate2}
\end{eqnarray}
where the first term corresponds to diagrams with two 
raritrons in the final state and the second with one. 
Notice that in both cases the temperature dependence differs from that in processes dominated by $\alpha_1$ discussed in the previous subsection. Indeed, because of the steep temperature dependence for two raritrons in the final state, the relic density from this term
will be sensitive to the maximum temperature, $T_{\rm max}$ attained 
during reheating, whereas the term stemming from single
raritron production will be sensitive only to $T_{\rm RH}$ \cite{Garcia:2017tuj}.

Following the procedure detailed in~\cite{Garcia:2017tuj}, we obtain the number density of dark matter at $T_{\rm RH}$, which takes the following form,
\begin{eqnarray}
n(T_{\rm RH})& = & \sqrt{\frac{10}{g_{\rm RH}}} \frac{0.995 \alpha_2^4 T_{\rm RH}^{10}}{\pi M_P^3 m_{3/2}^4} \ln\left(\frac{T_{\rm max}}{T_{\rm RH}}\right)   \nonumber \\ 
&& + \sqrt{\frac{10}{g_{\rm RH}}} \frac{0.011 \alpha_2^2 T_{\rm RH}^{6}}{\pi M_P m_{3/2}^2}\,,
\label{ntrh2}
\end{eqnarray}
where the ratio of $T_{\rm max}$ to $T_{\rm RH}$, assuming instantaneous thermalization, is given by \cite{Ellis:2015jpg} 
\beq
\frac{T_{\rm max}}{T_{\rm RH}} \simeq 0.5\left(\frac{m_{\Phi}}{\Gamma_{\Phi}}\right)^{1/4}\,, 
\eeq
From this we can 
calculate the present relic abundance at temperature $T_0$:
\begin{align}
\Omega_{3/2} h^2
 \;\simeq\; 0.1 &\left(\frac{\alpha_2}{6.2 \times10^{-9}}\right)^{2}\left(\frac{427 / 4}{g_{\text {RH }}}\right)^{3 / 2}  \nonumber \\
& \times \left(\frac{T_{\text {RH }}}{10^{10} \, \mathrm{GeV}}\right)^{3} \left(\frac{ 1\,\mathrm{GeV}}{m_{3 / 2}}\right)\, .
\label{Eq:omegabis2}
\end{align}
Note that this expression corresponds to that obtained from scatterings with a single raritron in the final state. Indeed, under the constraint that the raritron abundance saturates the observed dark matter relic density, single raritron production dominates over double raritron production for any phenomenologically sensible values of $T_{\rm RH}$ and $m_{3/2}$ (see Sec.~\ref{sec:results}). 

When thermalization is not assumed to occur instantaneously, the previous results can be significantly changed, given that the two raritron production cross section scales with the sixth power of the scatterer energy, that is $\sigma\sim s^3$, or $n=6$ in the classification of~\cite{Garcia:2017tuj}. In this case, the number density of non-thermally generated raritrons can be found to be given by
\beq
n(T_{\rm RH}) \;\simeq\; \left(\frac{5\pi^2 g_{\rm RH}}{72}\right)^{17/10} \frac{\alpha_2^4 m_{\Phi}^{24/5} T_{\rm RH}^{34/5}\mathcal{B}_2}{32400\pi \alpha_{\rm SM}^{16/5} m_{3/2}^4 M_P^{23/5}}\,.
\eeq
Here
\beq
\mathcal{B}_2 \;\equiv\; {\rm Br}_L + \frac{1}{12}{\rm Br}_{H}\,,
\eeq
where ${\rm Br}_L$ (${\rm Br}_{H}$) denotes the inflaton branching ratio to leptons (to Higgs). The corresponding non-thermal closure fraction can be written as follows,
\begin{align} \notag
\Omega_{3/2}h^2 \;\simeq\; &0.1 \left(\frac{\alpha_2}{5.9\times10^{-8}}\right)^4 \left(\frac{0.030}{\alpha_{\rm SM}}\right)^{16/5} \left(\frac{g_{\rm RH}}{427/4}\right)^{7/10} \\ \notag
& \times \left(\frac{1\,{\rm GeV}}{m_{3/2}}\right)^3 \left(\frac{m_{\Phi}}{3\times 10^{13}\,{\rm GeV}}\right)^{24/5}\\ 
& \times  \left(\frac{T_{\rm RH}}{10^{10}\,{\rm GeV}}\right)^{19/5} \mathcal{B}_2\,.
\label{nontherm2}
\end{align}
Comparing this expression with Eq.~(\ref{Eq:omegabis2}), we obtain that for
\beq
\left(\frac{m_{3/2}}{1\,{\rm GeV}}\right)\left(\frac{T_{\rm RH}}{10^7\,{\rm GeV}}\right)^{11/5} \;\lesssim\; 4\mathcal{B}_2\,,
\eeq
it is the non-thermal raritron population that dominates the dark matter energy budget. As in the case of the $\alpha_1$ coupling, in the discussion that follows we will omit this production mechanism due to its dependence on the inflationary model. 

In the previous subsection, we considered the production 
of raritrons from inflaton decay. We assumed that 
the inflaton was coupled only to the right-handed neutrino. 
In that case, raritron production from decay is possible only at 
two loops, and we do not consider that here. 
In principle one can couple the inflaton directly to the 
Standard model, but raritron production 
would be highly model dependent on the inflaton-SM coupling.
Therefore in the case of the $\alpha_2$ coupling, 
we do not consider raritron production from inflaton decay.

\section{Results and analysis}\label{sec:results}

\subsection{Results when $\alpha_1$ dominates}

In the preceding analysis, we derived the raritron
lifetime and density in terms of the Dirac coupling $y$,
the right-handed neutrino mass, $M_R$, and the light neutrino
mass, $m_1$, though these are related through Eq.~(\ref{Eq:m1}). 
In addition, there is  
an absolute theoretical 
limit on $y$ ($y \lesssim \sqrt{4 \pi}$) from perturbativity and
an experimental
cosmological constraint on the sum of the light neutrino masses which force $m_1 \lesssim 0.15$ eV \cite{sumnu}.
In other words, for a given $M_R$, the upper bound on $m_1$ implies an upper bound on $y$, and as a consequence a lower bound on the raritron lifetime 
and upper bound on its relic abundance. For example,
\beq
m_1 \lesssim 0.15~  \mrm{eV} ~~\Rightarrow ~~ y\lesssim  0.7 \sqrt{\frac{M_R}{10^{14}~{\rm GeV}}} \, .
\eeq
We can then express the lifetime constraints (\ref{Eq:lifetime})
as a function of $m_1$
\beq
\Gamma_{3/2}=\Gamma_{3/2}^{2b} + \Gamma_{3/2}^{3b}=
\frac{\alpha_1^2}{4 \pi} \frac{m_1 m_{3/2}^3}{M_P^2 M_R}\left[1 + \frac{m_{3/2}^2}{60 \pi^2v^2}  \right] \, .
\eeq
Given $m_{3/2}$ and $\alpha_1$,
limits from the dark matter lifetime (\ref{Eq:lifetime})
give us a lower bound on $M_R$ (we fix $m_1=0.15$ eV to be specific). For dark matter production from scattering, we can use the lower bound on $M_R$
to find the reheating temperature $T_{\rm RH}$
necessary to obtain the correct relic abundance in Eq.~(\ref{Eq:omegabis}). For dark matter produced from decay, the bound on $M_R$ is not needed.  Note that when we saturate the bound on $m_1$ we also have an upper limit on $M_R$ from the perturbativity of $y < \sqrt{4 \pi}$, which is $M_R \lesssim 2.5 \times 10^{15}$ GeV.

\subsubsection{Dark matter production from scattering}

\begin{figure}[t]
\centering
\includegraphics[width=\columnwidth]{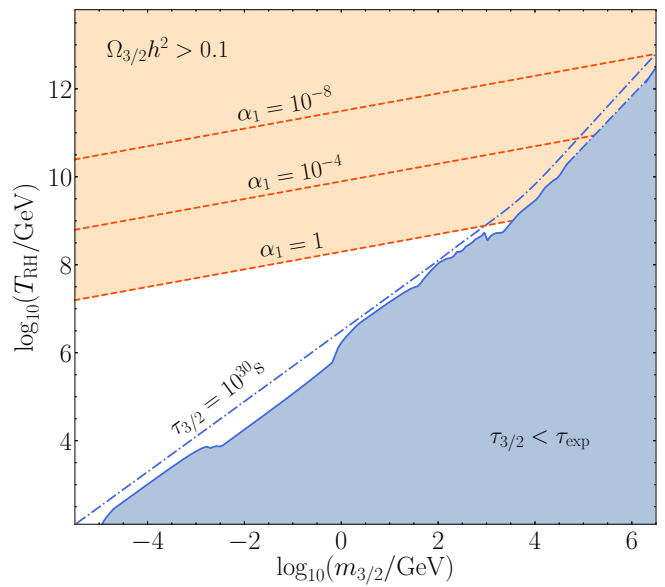}
\caption{\em \small The ($m_{3/2}$, $T_{\rm RH}$) plane with astrophysical constraints on the lifetime from $\gamma$-ray observations and {\em Planck} constraints on the relic abundance for different values of $\alpha_1$ ($10^{-8}$, $10^{-4}$ and 1) and $m_1=0.15~\mrm{eV}$. See the text for details. 
}
\label{Fig:results}
\end{figure}

We consider first the case where dark matter is produced
exclusively through scatterings during reheating. 
That is, we assume that the direct production from inflaton decay is negligible. We show in Fig.~\ref{Fig:results} the available parameter space in the ($m_{3/2},T_{\rm RH}$) plane. In the lower 
right portion of the plane, the raritron lifetime is too short
when compared with experimental constraints. 
Due to the large range of dark matter masses we apply constraints from several experiments: XMM-Newton observations of M31 \cite{Boyarsky:2007ay} at the keV scale, SPI, INTEGRAL and COMPTEL observations \cite{Boyarsky:2007ge,Yuksel:2007dr} at the MeV scale, the latest limits from FERMI-LAT at the GeV scale \cite{Ackermann:2015lka}, and HESS above the TeV scale \cite{Abramowski:2013ax} (see also \cite{Lattanzi:2013uza}). Note that the limits on the dark matter lifetime given by the collaborations correspond to a specific final state. A complete study taking into account the exact shape of the spectrum is
beyond the scope of our work, and is not necessary considering the large dependence of the relic abundance on the reheating temperature. 

To obtain the limit on the lifetime in the 
($m_{3/2},T_{\rm RH}$) plane, we first fix the value of
$\tau_{3/2}$ at the experimental limit from Eq.~(\ref{Eq:lifetime}) for each value of $m_{3/2}$. This determines the combination $M_R/\alpha_1 y$. Then from Eq.~(\ref{Eq:omegabis}), 
we can determine the value of $T_{\rm RH}$ needed to obtain
$\Omega_{3/2} h^2 \simeq 0.1$. This procedure determines the blue line
in Fig.~\ref{Fig:results}. For lower masses ($< 10$ TeV), we use $\gamma$-ray limits, whereas for higher masses ($> 1$ PeV) we use neutrino limits and the dot-dashed portion of the line in between is an extrapolation. In the shaded region below this line,
we continue to fix $\Omega_{3/2} h^2 \simeq 0.1$, but to do so 
at lower $T_{\rm RH}$ requires lower values of $M_R/\alpha_1 y$
and hence lifetimes below the experimental limit. Conversely,
in this shaded region, satisfying the lifetime limit would
imply an insufficient relic density (though this cannot be excluded). For reference we also plot in Fig.~\ref{Fig:results} the line corresponding to a projected sensitivity corresponding to a lifetime of $\tau_{3/2}= 10^{30}~{\rm s}$ which is similar to the present experimental limits.
This line can be determined from the substitution of $\tau_{3/2}$
into $\Omega_{3/2} h^2$ giving
\beq
\Omega_{3/2} h^2 \simeq 0.1 \times \left( \frac{3 \times 10^{31} \, {\rm s}}{\tau_{3/2}^{\mrm{2b}}} \right)
\left( \frac{10^4}{m_{3/2}} \right)^4
\left( \frac{T_{\rm RH}}{10^{10}} \right)^5 \, ,
\label{combsc}
\eeq
resulting in a slope of 4/5 (in the logs) for $T_{\rm RH}$ vs $m_{3/2}$. 
A change of slope in this line occurs for $m_{3/2} = 2\sqrt{15} \pi v \simeq 6$ TeV corresponding to the point when the three-body and two-body decay rates are equal. At higher masses, using $\tau_{3/2}^{\mrm{3b}}$ we have, 
\beq
\Omega_{3/2} h^2 \simeq 0.1 \times \left( \frac{10^{31} \, {\rm s}}{\tau_{3/2}^{\mrm{3b}}} \right)
\left( \frac{10^4}{m_{3/2}} \right)^6
\left( \frac{T_{\rm RH}}{10^{10}} \right)^5 \, ,
\label{combsc3}
\eeq
which results in a slope of 6/5. 

In the upper left portion of the ($m_{3/2},T_{\rm RH}$) plane
we fix the value of $M_R = 2.5 \times 10^{15}$ GeV at its perturbative limit from $y < \sqrt{4 \pi}$. In this region, above the blue line, 
the lifetime is always longer than the experimental limit. 
Assuming $m_1 = 0.15$ eV,  we show three contours with fixed $\alpha_1$ as indicated and $\Omega_{3/2} h^2 = 0.1$. For each value
of $\alpha_1$, the shaded region above the line (at higher $T_{\rm RH}$) would have an excessive raritron density.
We see immediately that raritron masses from about a keV to a PeV
are all allowed for reasonable reheat temperatures $T_{\rm RH} \gtrsim 10^6$ GeV.

\begin{figure}[t]
\centering
\includegraphics[width=\columnwidth]{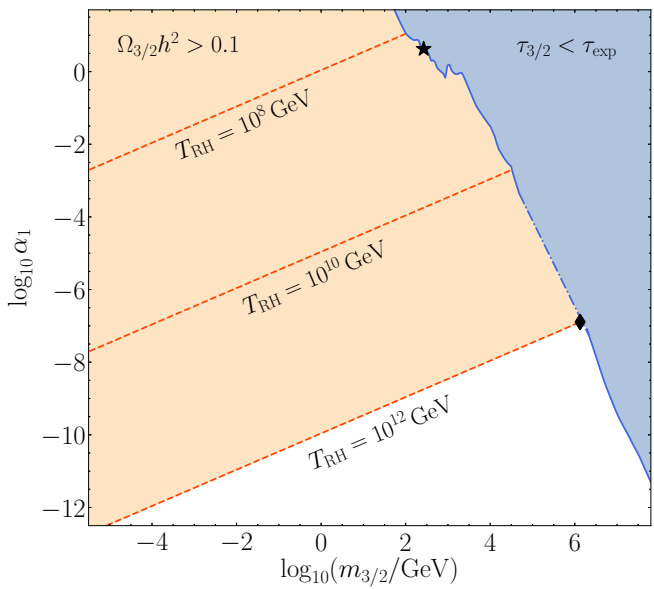}
\caption{\em \small The ($m_{3/2}$, $\alpha_1$) plane with astrophysical constraints on the lifetime from the $\gamma$-ray observations and {\em Planck} constraints on the relic abundance produced by scattering for $M_R=10^{14} \, \rm{GeV}$ , $m_1=0.15~\mrm{eV}$ and different values of $T_{\rm RH}$ ($10^{8}$, $10^{10}$, and $10^{12} \, \rm{GeV}$). See the text for details.
}
\label{Fig:resultsalpha}
\end{figure}

It is also useful to consider the allowed parameter space in the ($m_{3/2}$, $\alpha_1$)
plane. We show in Fig.~\ref{Fig:resultsalpha} the region allowed for different values of the reheating temperature as indicated. 
The curves and shadings are as in the previous figure, however,
we now fix both $m_1 = 0.15$ eV and $M_R = 10^{14}$ GeV everywhere across the plane. In this case, the lifetime limit shown by the blue curve can be viewed as a function of $m_{3/2}$ and $\alpha_1$ and should be close to a line with log slope of $-3/2$ for low $m_{3/2}$ and $-5/2$ for larger masses when the three-body decay dominates. 
As discussed above, we find that values of $\alpha_1$ of order 1 are allowed for relatively low reheating temperatures ($T_{\rm RH} \simeq 10^8~{\rm GeV}$) whereas higher reheating temperatures of order $10^{12}$ necessitate $\alpha_1 \lesssim 10^{-8}$ to avoid an overabundance of dark matter. 

\subsubsection{Including the inflaton decay}

Reheating is the result of inflaton decay to SM particles.
If $M_R < m_\Phi$, there will be tree-level diagrams which
lead to reheating and raritron production. 
When $M_R > m_\Phi$,
there may be a direct coupling between the inflaton and the SM
(characterized by the coupling $y_\Phi > y_\nu$ in Eq.~(\ref{infdec})) 
or through loops with $y_\Phi = y_\nu$ as in Fig.~\ref{brl1} and discussed earlier. 
Even if there is no direct coupling between the inflaton and raritron, raritron production through loops is possible as in Fig.~\ref{brl2}. 
Unless inflaton decay to dark matter is highly suppressed, once a direct decay channel is open even through loops, it can easily dominate the dark matter production \cite{Kaneta:2019zgw}. We have seen this effect for the specific case of raritron dark matter in the preceding section.

\begin{figure}[t]
\setlength\belowcaptionskip{-0.5\baselineskip}
\centering
\includegraphics[width=\columnwidth]{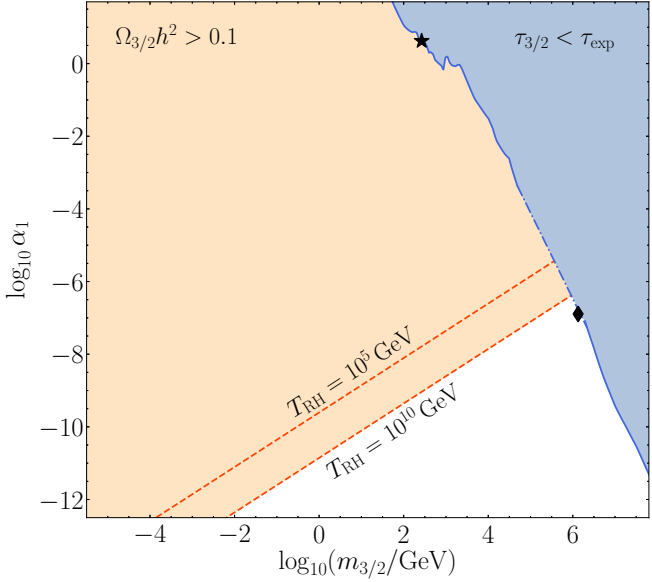}
\caption{\em \small The ($m_{3/2}$, $\alpha_1$) plane with astrophysical constraints on the lifetime from the $\gamma$-ray observations  and {\em Planck} constraints on the relic abundance produced by inflaton decay for $M_R=10^{14}$ GeV, $m_1=0.15~\mrm{eV}$ and two values of $T_{\rm RH}$ ($10^{5}$ and $10^{10}$ GeV). See the text for details. 
}
\label{Fig:resultsbr}
\end{figure}

To get an idea of the relevant parameter values,
we rewrite Eq.~(\ref{Eq:omegabr}) (ignoring the logs) 
with $\Omega_{3/2} h^2|_{\rm decay} \simeq 0.1$ as
\beq
m_{3/2}\simeq 4 \times 10^{14}
\alpha_1^{\frac{4}{3}} \left( \frac{0.15~{\rm eV}}{m_1} \right)^{2/3} \left( \frac{T_{\rm RH}}{10^{10}~\mrm{GeV}}\right)^{\frac{1}{3}}
\rm{GeV} \, .
\eeq
Using this value for $m_{3/2}$ in the lifetime in Eq.~(\ref{Eq:lifetime}) (using the three-body decay as an example), we find for $\tau_{3/2} \gtrsim 10^{30}~\rm{s}$
\begin{align} \notag
\alpha_1 \;\lesssim\; 2 &\times 10^{-7}
\left(\frac{M_R}{10^{14}~{\rm GeV}} \right)^{\frac{3}{26}} 
\left( \frac{m_1}{0.15~{\rm eV}} \right)^{7/26}\\
&\times
\left(\frac{10^{10}\,{\rm GeV}}{T_{\rm RH}}\right)^{\frac{5}{26}}\, .
\label{combdec}
\end{align}
The relevant parameter space in the ($m_{3/2}$, $\alpha_1$)
plane is shown in Fig.~\ref{Fig:resultsbr}.
Since the blue line is determined solely from the limit on the raritron lifetime, it is independent of the production mechanism 
and is the same as in Fig.~\ref{Fig:resultsalpha}.
Comparing Fig.~\ref{Fig:resultsbr} with Fig.~\ref{Fig:resultsalpha}, we see clearly that the production of dark matter through inflaton decay is much more copious and the parameter space is much more constrained. The relic abundance necessitates much lower values of $\alpha_1$ to avoid overabundance, and the result is much less dependent on $T_{\rm RH}$ as one can see comparing Eqs.~(\ref{Eq:omegabis}) and (\ref{combsc}), where
$\Omega_{3/2} h^2$ depends on $T_{\rm RH}^5$ in the scattering case, compared to the $T_{\rm RH}$ inflaton decay process. This feature is also clearly illustrated in Fig.~\ref{Fig:resultsbr}, where we show two lines producing the correct relic abundance with $T_{\rm RH}=10^5$ and $10^{10}$ GeV.

\subsection{Results when $\alpha_2$ dominates}\label{sec:alpha2phen}

\begin{figure}[t]
\centering
\includegraphics[width=\columnwidth]{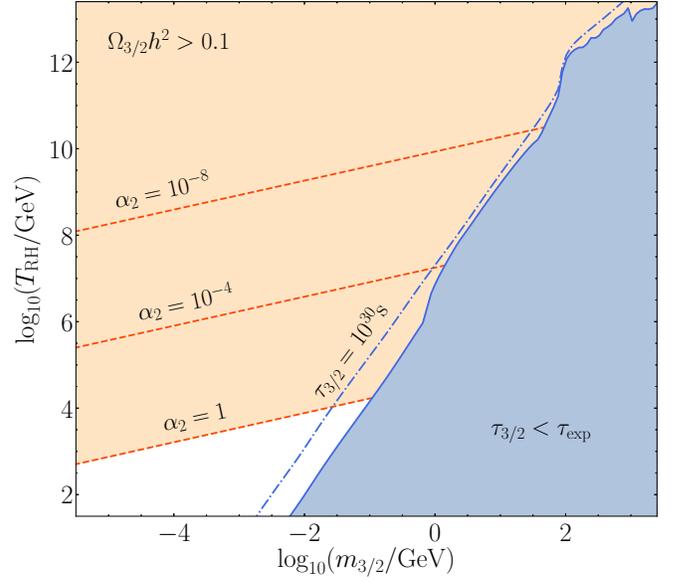}
\caption{\em \small As in Fig.~\ref{Fig:results}, the ($m_{3/2}$, $T_{\rm RH}$) plane with astrophysical constraints on the lifetime from $\gamma$-ray observations and {\em Planck} constraints on the relic abundance for different values of $\alpha_2$ ($10^{-8}$, $10^{-4}$ and 1).
}
\label{Fig:results2}
\end{figure}

We can now repeat the previous analysis when the 
raritron is coupled to the SM through $\alpha_2$. 
We first show the results in the ($m_{3/2},T_{\rm RH}$) plane in Fig.~\ref{Fig:results2} produced in the same manner as in the previous subsection. Due to the model dependence of the abundance of produced raritrons from pre-thermal scatterings, encoded in the effective branching ratio $\mathcal{B}_2$, we limit ourselves here to ratiron production in thermal equilibrium, i.e.~Eq.~(\ref{Eq:omegabis2}). We note that while there is not a direct
raritron decay channel with a photon in the final state, photons
will be produced. However, in the absence of a simulation 
of the full decay chain, we show the same experimental limits used in the previous subsection. We also show for reference, the curve
corresponding to a lifetime of $10^{30}$ s. 
The experimental constraints demanding a sufficiently long lifetime and suitable relic density push the allowed mass range 
to $m_{3/2} \lesssim 100$ GeV when $T_{\rm RH} \sim 10^{12}$ GeV,
and $m_{3/2} \lesssim 1$ GeV when $T_{\rm RH} \sim 10^{8}$ GeV.
Note that our numerical results include the running of the gauge and Yukawa couplings to the appropriate temperature scale, 
though not explicitly written in the analytic formulas. 

\begin{figure}[t]
\setlength\belowcaptionskip{-0.5\baselineskip}
\centering
\includegraphics[width=\columnwidth]{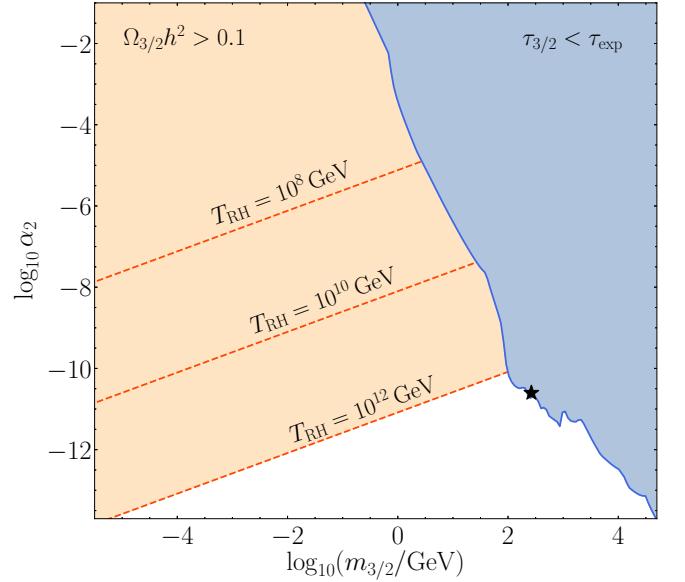}
\caption{\em \small As in Fig.~\ref{Fig:resultsalpha}, the ($m_{3/2}$, $\alpha_2$) plane with astrophysical constraints on the lifetime from the $\gamma$-ray observations and {\em Planck} constraints on the relic abundance produced by scattering for different values of $T_{\rm RH}$ ($10^{8}$, $10^{10}$, and $10^{12} \, \rm{GeV}$). 
}
\label{Fig:resultsalpha2}
\end{figure}

As one can see from comparing Figs.~\ref{Fig:results} and \ref{Fig:results2}, for a given reheat temperature,
and relic density, we need $\alpha_2 \ll \alpha_1$ as noted earlier. For example, for $m_{3/2} \sim 10$ MeV, and $T_{\rm RH} \sim 10^{8}$ GeV, we find $\alpha_1 \sim 10^{-2}$, whereas, $\alpha_2 \sim 10^{-7}$. 
Similarly, we show the results in the ($m_{3/2}$, $\alpha_2$) plane in Fig.~\ref{Fig:resultsalpha2}. We see that unless $\alpha_2$ is very small ($< 10^{-11}$), $m_{3/2} < 100$ GeV, and the upper limit on the raritron mass decreases with increasing $\alpha_2$.

We also see by comparing Figs.~\ref{Fig:resultsalpha}
and \ref{Fig:resultsalpha2}, that when the reheat temperature
is rather large ($T_{\rm RH} \sim 10^{12}$ GeV), and $m_{3/2} \sim 10$ MeV, the values of $\alpha_{1,2}$ needed to produce
$\Omega h^2 = 0.12$ are similar though quite small. The required
difference in the couplings is much greater at smaller $T_{\rm RH}$.

\section{Signatures}

Having established the viable parameter space for the raritron dark matter model, we now discuss in more detail the 
possible experimental signatures for such a model.
The two-body decay mode shown in Fig.~\ref{Fig:feynman1}
will produce a monochromatic photon and neutrino.\footnote{A signal of this type was termed a ``double smoking-gun'' in \cite{Aisati:2015ova}. Spin-3/2 fields were not included in  their study, nor in \cite{ProfumoPlanck}, and they did not try to produce cosmologically viable scenarios.} If $m_{3/2}> m_Z$, there is also a two-body final state $Z + \nu_1$. These decay channels are easily observable at detectors and the signal will give us 1) the mass of the dark matter (from the position of the signal in the spectrum) and 2) the lifetime (from the strength of the signal). On top of the monochromatic signal, there will also be a continuous spectrum due to the three-body channels
which dominate at higher raritron masses. If the raritron is 
produced mainly through scattering, the signal
can be translated to the reheating temperature needed to obtain the right relic abundance using Eq.~(\ref{combsc}).
For example, a GeV gamma ray observed by FERMI 
with a signal strength corresponding to a lifetime of $10^{30}$ s
would imply a reheating temperature of $\simeq 3\times 10^6$ GeV. In this case, the temperature is {\it independent} of the parameter $\alpha_1$. In contrast, if the production were dominated by inflaton decays, some information on the combination of 
$\alpha_1$ and $T_{\rm RH}$ could be ascertained.

As an exercise, we reanalyze one of the most popular recent ``signals'': the 130 GeV line observed by the FERMI satellite in 2012 \cite{Ackermann:2012qk}. This is a monochromatic signal that could be fit with a dark matter mass $m_{3/2}\simeq 260$ GeV and a lifetime $\Gamma_{3/2}\simeq 10^{29}~{\rm s}$\cite{Kyae:2012vi}, for a Navarro, Frenk, and White profile~\cite{Navarro:1995iw}.\footnote{The dependence on the dark matter distribution for decaying dark matter being proportional to its density $\rho$ (versus $\rho^2$) is much weaker than for annihilating dark matter.} Interestingly, this kind of signal could correspond to a spin-$\frac32$ dark matter decay. For scattering-dominated production,
we can use Eq.~(\ref{combsc}) to determine the reheat temperature,
$T_{\rm RH}\simeq 2 \times 10^8$ GeV. From the lifetime, we can 
also determine the combination $(\alpha_1 y/M_R)^2$ or $\alpha_1^2/M_R$ upon fixing the light neutrino mass, $m_1$. 
We find, $\alpha_1^2/M_R  = 2 \times 10^{-13}$ GeV$^{-1}$ or
$\alpha_1 \approx 4$ for $M_R = 10^{14}$ GeV.  
Note that for this value of $\alpha_1$, when inflaton decay
is the dominant production mode, the reheating temperature
must be extremely (and unphysically) low. Thus
not only would we determine the reheat temperature and $\alpha_1$,
but we would also know that inflaton decay does not play a role in dark matter production. This position of this example in the ($m_{3/2}$, $\alpha_1$) plane is 
illustrated in Figs.~\ref{Fig:resultsalpha} and \ref{Fig:resultsbr} by a star.
Such a signal, if observed  below $m_H$, could be correlated with a similar monochromatic signal from neutrino detectors like ANTARES or IceCube\footnote{
The latter being more sensitive to track events than the ANTARES telescope.}. 
Even if dedicated neutrino line searches have not been yet performed by the IceCube Collaboration, its sensitivity has been calculated in \cite{ElAisati:2017ppn} and should probe a lifetime $\Psi_\mu \rightarrow A_\mu + \nu$ of $\simeq 10^{29}~{\rm s}$.

Next, we repeat the exercise for the PeV neutrino signal observed by IceCube \cite{Aartsen:2014gkd}. 
There were some attempts to explain these events from a dark matter perspective (see \cite{Dudas:2014bca} for instance) but it was difficult to reconcile the signal with the correct relic abundance. 
The number of events expected by IceCube is \cite{Gandhi:1998ri,Dudas:2014bca}
\beq
\Gamma_{\rm events} = 1.5 \times 10^{57} \eta_E f_{\rm astro} \frac{\Gamma_{3/2}}{m_{3/2}^{0.637}}~\rm{years^{-1}},
\eeq
where $\eta_E\sim 0.4$ is defined from the fiducial volume $V_{\rm fid} = \eta_E V$ and $f_{\rm astro}\sim 1$ corresponds to the astrophysical uncertainty in the local distribution of the dark mater halo. The mass and widths are expressed in GeV. A rate of one PeV event per year gives us 
$\Gamma_{3/2} \simeq 10^{-53}~\mrm{GeV}$, corresponding to 
$\tau_{3/2} \simeq 6 \times 10^{28}$ s.\footnote{This is 
similar to what was obtained in \cite{Feldstein:2013kka},
namely, $\tau_{3/2} \simeq 1.9 \times 10^{28}~{\rm s}$. }
Using this lifetime, with $m_{3/2} = 1$ PeV, we can again
determine the value of $\alpha_1$ now from the three-body decay rate which is dominant, $\alpha_1 \simeq 10^{-7}$. The reheating temperature in this case can be obtained from Eq.~(\ref{combsc3}) and we find
$T_{\rm RH} \simeq 9 \times 10^{11}$ GeV, when raritron production is due to scattering. When production is due to inflaton decay, 
we can use Eq.~(\ref{combdec}) and find, $T_{\rm RH} \simeq 4\times10^{11}$~GeV. The position of this example is displayed in 
 Figs.~\ref{Fig:resultsalpha} and \ref{Fig:resultsbr} by the black diamond.  Both the scattering production and the inflaton decay process scenarios are compatible correct relic abundance {\it and} the IceCube PeV monochromatic signals.

\section{Conclusion}

We have shown that a metastable spin-$\frac{3}{2}$ particle can be a suitable dark matter candidate through the introduction of a minimal (Planck-suppressed) coupling, $\alpha_1$, to a right-handed neutrino. Surprisingly, the parameter space needed to generate a sufficiently long lifetime is perfectly compatible with both the 
astrophysical constraints from $\gamma$-ray and neutrino experiments as well as the cosmological determination of the dark matter density
. 
Our results are summarized in Figs.~\ref{Fig:resultsalpha} and \ref{Fig:resultsbr} where we display the allowed region in the 
($m_{3/2}, \alpha_1$) plane. We considered both 
the production of dark matter from the thermal bath produced during reheating, and production directly from inflaton decay. We also have shown that smoking-gun signals are expected from such couplings, in the form of a monochromatic neutrino {\it and/or} a monochromatic gamma-ray
line.

In addition, we considered a second possible gauge invariant coupling, $\alpha_2$ of the raritron to the SM Higgs and lepton doublet. The requirement of a sufficiently long lifetime and 
correct relic abundance, restricts the raritron mass range to lower values and lower coupling as seen in Figs.~\ref{Fig:results2} and \ref{Fig:resultsalpha2}.

We have also illustrated, as examples, the points in the parameter space that could explain the gamma-ray signal observed by the FERMI telescope, or PeV neutrinos observed by IceCube that can be combined with the recent ANITA analysis \cite{ANITA}. 
 Moreover, it was shown in \cite{benakli2} that spin-$\frac{3}{2}$ particles can have an impact on the form of gravitational waves produced during reheating that could be observable in future ultra-high frequency detectors.

\vskip.1in
{\bf Acknowledgments:}
\noindent 
The authors want to thank especially Kunio Kaneta and Emilian Dudas for very  insightful
discussions. This work was supported in part by the France-U.S. PICS MicroDark.
The work of M.A.G.G.~was supported by the Spanish Agencia
Estatal de Investigaci\'on through the Grants No.~FPA2015-65929-P (MINECO/FEDER, UE) and No.~PGC2018095161-B-I00, IFT Centro de Excelencia Severo
Ochoa Grant No.~SEV-2016-0597, and Red Consolider MultiDark
Grant No.~FPA2017-90566-REDC. This project has received funding/support from the European Union's Horizon 2020 research and
innovation program under the Marie Skodowska-Curie Grant Agreements Elusives ITN No.~674896
and InvisiblesPlus RISE No.~690575. The work of K.A.O.~was supported in part by the U.S.~DOE Grant No.~DE-SC0011842 at the University of Minnesota.
\vspace{-0.2cm}
\appendix

\section{The Rarita-Schwinger Lagrangian}
\label{app:RS}

Rarita and Schwinger \cite{Rarita:1941mf} derived the Lagrangian
(\ref{Eq:rarita}) following the work of Fierz and Pauli \cite{Fierz:1939ix}. One can start with the hypothesis that a spin-$\frac{3}{2}$ particle should respect both the spin-$\frac{1}{2}$ Dirac equation {\it and} spin-1 divergence relation, namely,
\bea
&&
(i \gamma^\rho \partial_\rho - m_{3/2}) \Psi_\mu = 0
\label{Eq:raritabis}
\\
&&
\partial^\mu \Psi_\mu =0 .
\label{Eq:raritater}
\eea
By writing the field $\Psi$ in terms of its spin 
components and after a Clebsch-Gordan decomposition, we have
\bea
&&
\bar \Psi_\mu^{+\frac{3}{2}}= \Psi^{+\frac{1}{2}} \epsilon_\mu^{+1} ,
\nonumber
\\
&&
\Psi_\mu^{+\frac{1}{2}}= \frac{1}{\sqrt{3}}\Psi^{-\frac{1}{2}} \epsilon_\mu^{+1}
+\sqrt{\frac{2}{3}} \Psi^{+ \frac{1}{2}}\epsilon_\mu^0 ,
\nonumber
\\
&&
\Psi_\mu^{-\frac{1}{2}}= \frac{1}{\sqrt{3}}\Psi^{+\frac{1}{2}} \epsilon_\mu^{-1}
+\sqrt{\frac{2}{3}} \Psi^{-\frac{1}{2}}\epsilon_\mu^0 ,
\nonumber
\\
&&
\Psi_\mu^{-\frac{3}{2}}= \Psi^{-\frac{1}{2}} \epsilon_\mu^{-1} ,
\label{comp}
\eea
where $\Psi^{s_z}$ is a Dirac spinor of helicity $2s_z$, which is a solution of Eq.~(\ref{Eq:raritabis}), and
$\epsilon_\mu^\lambda$ is a vector polarization with spin projection $\lambda$ along the direction of the momentum, so that  $\partial^\mu \epsilon_\mu = 0$. See \cite{Christensen:2013aua} for a detailed solution.
One can show, using each of the components in Eq.~(\ref{comp}) by direct calculation and after a little algebra, that Eqs.~(\ref{Eq:raritabis}) and (\ref{Eq:raritater}) imply
\beq
\gamma^\mu \Psi_\mu=0.
\label{Eq:rarita4}
\eeq

We can construct a Lagrangian for a spin-$\frac{1}{2}$ field, whose Euler-Lagrange equation gives Eq.~(\ref{Eq:raritabis}), with terms such as
$\gamma^\mu \gamma^\nu$, $\gamma^\mu \partial^\nu$, or any combinations of that type, which are consistent with the relations (\ref{Eq:raritater}) and (\ref{Eq:rarita4}). Among the class of possible Lagrangians, the simplest one is
\bea
&&
{\cal L}_{3/2}^0=\Psi_\mu
\left(  
i g^{\mu \nu} \gamma^\rho \partial_\rho - m_{3/2} g^{\mu \nu}
\right.
\nonumber
\\
&&
\left.
-i \gamma^\mu \partial^\nu -i\gamma^\nu \partial^\mu 
+i \gamma^\mu \gamma^\rho \gamma^\nu \partial_\rho +m_{3/2}\gamma^\mu \gamma^\nu
\right) \Psi_\nu .
\label{rssimple}
\eea
Note that the coefficient of the last four terms in Eq.~(\ref{rssimple}) is arbitrary (e.g., \cite{Rarita:1941mf} included a factor of 1/3 in front of each of these terms). Equation (\ref{rssimple})
can be simplified to
\beq
\bar \Psi_\mu \left(i \gamma^{\mu \rho \nu} \partial_\rho + m_{3/2} \gamma^{\mu \nu} \right) \Psi_\nu 
\label{Eq:raritafinal}
\eeq
which is, up to a normalization factor, our Lagrangian in Eq.~(\ref{Eq:rarita}).

\section{Decay and scattering rates}
\label{app:rates}

In this appendix, we provide some relevant details concerning the computation of the dark matter decay rate. 

\subsection{Three-body decay formula}

The phase space integration for the three-body decay processes $\Psi_{\mu} \rightarrow A_{\mu} H \nu_1$ and $\Psi_{\mu} \rightarrow Z_{\mu} H \nu_1$ can be performed analytically if one disregards the small neutrino mass $m_1$. In this limit, the decay rates are given by Eqs.~(\ref{eq:3ba}) and (\ref{eq:3bb}), where the threshold function is given by the following expression,
\begin{widetext}
\begin{align} \notag
g(x,y) \;=\; \Bigg[ &\left(1+\frac{y^2}{4}\right) \left(1-y^2\right)^3 +\frac{113}{8} x^2 \left(1-\frac{119 y^2}{339} -\frac{29 y^4}{339} -\frac{131 y^6}{339}\right) +\frac{59}{24} x^4 \left(1 +\frac{6 y^2}{59} -\frac{51 y^4}{59}\right)\\ \notag
& -\frac{1}{24} x^6 \left(1-9 y^2\right) -\frac{x^8}{24} \Bigg] \xi -5 x^2 y^6 \left(2 x^2+y^2\right) \ln \left| 2 x y \left(\xi -x^2-y^2+1\right) \right|\\
& +\frac{5}{2} x^2 \left(4 x^2-4 y^2+3\right) \ln \left| \frac{x^4-x^2 \left(\xi +2 y^2+1\right)+y^2 \left(\xi+y^2-1\right)}{\xi-x^2-y^2+1} \right| \, ,
\end{align}
where
\begin{align}
\xi \;=\; \sqrt{x^4-2 x^2 \left(y^2+1\right)+\left(1-y^2\right)^2}\,.
\end{align}
\end{widetext}

For the three-body decay processes $\Psi_{\mu} \rightarrow \nu_{\ell} \bar{f} f$, we find
\begin{widetext}
\begin{align} \notag
 g_1(x) =\; &  30 \left(14 x^{8}-24 x^{6}+18 x^{4}-8 x^{2}+3 \right) x^{4} \times \ln \left[\frac{4 x^{4}}{(1-\sqrt{1-4 x^{2}})\left(1-\sqrt{1-4 x^{2}}+(\sqrt{1-4 x^{2}}-3) x^{2}\right)}\right] \\
&+  (x-1)(x+1)\left(420 x^{8}-230 x^{6}+204 x^{4}+33 x^{2}-1\right) \sqrt{1-4 x^{2}},
\end{align}
and
\begin{align} \notag
 g_2(x) =\; & 20 x^{4} \ln \left[\frac{4 x^{4}+4(\sqrt{1-4 x^{2}}-2) x^{2}-2 \sqrt{1-4 x^{2}}+2}{4 x^{4}}\right] \left(c_{V}^{2}\left(x^{6}+3 x^{4}-3 x^{2}+1\right)-c_{A}^{2}\left(-x^{6}+3 x^{4}-3 x^{2}+2\right)\right) \\
& + \sqrt{1-4 x^{2}}\left[c_{V}^{2}\left(-20 x^{8}-\frac{190 x^{6}}{3}+\frac{148 x^{4}}{3}-3 x^{2}+1\right) +c_{A}^{2}\left(-20 x^{8}+\frac{170 x^{6}}{3}-\frac{152 x^{4}}{3}-23 x^{2}+1\right) \right]\,.
\end{align}
\end{widetext}

\subsection{Scattering amplitudes}

The amplitudes for the scattering processes contributing to dark matter production from the thermal bath, for the case with coupling $\alpha_1$, can be written as
\begin{align} \label{eq:MHnupsiB} \displaybreak[0]
    |\mathcal{M}|^2_{H \nu_1 \rightarrow \Psi_{\mu} B} &= -\frac{8}{3} \frac{\alpha_1^2 y^2}{m_{3/2}^2 M_P^2} \left(\frac{s}{s - M_R^2}\right)^{2} \left(M_R^2 t + su \right) \,,\\ \displaybreak[0]
    |\mathcal{M}|^2_{H B \rightarrow \Psi_{\mu} \nu_1} &= \frac{8}{3} \frac{\alpha_1^2 y^2}{m_{3/2}^2 M_P^2} \left(\frac{t}{t - M_R^2}\right)^{2} \left(M_R^2 s + ut \right) \,,\\ \displaybreak[0]
    |\mathcal{M}|^2_{B \nu_1 \rightarrow \Psi_{\mu} H} &= -\frac{8}{3} \frac{\alpha_1^2 y^2}{m_{3/2}^2 M_P^2} \left(\frac{t}{t - M_R^2}\right)^{2} \left(M_R^2 u + st \right) \, .
\end{align}
In the limit of $M_{R} \gg m_{3/2}$, we find 
\begin{align} \displaybreak[0]
    |\mathcal{M}|^2_{H \nu_1 \rightarrow \Psi_{\mu} B} \;&=\; -\frac{8}{3} \frac{\alpha_1^2 y^2}{m_{3/2}^2 M_R^2 M_P^2} s^2t\,,\\ \displaybreak[0]
    |\mathcal{M}|^2_{H B \rightarrow \Psi_{\mu} \nu_1} \;&=\; \frac{8}{3} \frac{\alpha_1^2 y^2}{m_{3/2}^2 M_R^2 M_P^2} st^2\,,\\
    |\mathcal{M}|^2_{B \nu_1 \rightarrow \Psi_{\mu} H} \;&=\; -\frac{8}{3} \frac{\alpha_1^2 y^2}{m_{3/2}^2 M_R^2 M_P^2} ut^2\,.
\end{align}

Similarly, for the interactions mediated by the coupling $\alpha_2$, assuming $T \gg m_{3/2}$, we find
\begin{equation}
|\mathcal{M}|^2_{Lf \rightarrow \Psi_{\mu}f'} = \frac{3}{8} \frac{\alpha_2^2 m_t^{2}}{v^2 m_{3/2}^2 M_P^2} \frac{(t - m_{3/2}^2)^3}{t}\,,
\end{equation}
\begin{equation}
|\mathcal{M}|^2_{ff' \rightarrow \Psi_{\mu}L} = \frac{3}{8} \frac{\alpha_2^2 m_t^{2}}{v^2 m_{3/2}^2 M_P^2} \frac{(s - m_{3/2}^2)^3}{s}\,,
\end{equation}
\begin{equation}
|\mathcal{M}|^2_{LW \rightarrow \Psi_{\mu}H} = \frac{3}{32} \frac{\alpha_2^2 g^2}{m_{3/2}^2 M_P^2} (s - 2t)(s - t)\,,
\end{equation}
\begin{equation}
|\mathcal{M}|^2_{LB \rightarrow \Psi_{\mu}H} = \frac{1}{32} \frac{\alpha_2^2 g'^2}{m_{3/2}^2 M_P^2} (s+t)(s+2t)\,,
\end{equation}
\begin{equation}
|\mathcal{M}|^2_{LH \rightarrow \Psi_{\mu}W} = \frac{3}{16} \frac{\alpha_2^2 g^2}{m_{3/2}^2 M_P^2} (s + 2t)(s + 3t)\,,
\end{equation}
\begin{equation}
|\mathcal{M}|^2_{LH \rightarrow \Psi_{\mu}B} = \frac{1}{16} \frac{\alpha_2^2 g'^2}{m_{3/2}^2 M_P^2} s(s-t)\,,
\end{equation}

\begin{equation}
|\mathcal{M}|^2_{HW \rightarrow \Psi_{\mu}L} = \frac{3}{16} \frac{\alpha_2^2 g^2}{m_{3/2}^2 M_P^2} (2s^2 - 3 st + t^2)\,,
\end{equation}

\begin{equation}
|\mathcal{M}|^2_{HB \rightarrow \Psi_{\mu}L} = \frac{1}{16} \frac{\alpha_2^2 g'^2}{m_{3/2}^2 M_P^2} (s +t)(2s + t)\,,
\end{equation}

\begin{equation}
    |\mathcal{M}|^2_{H \ell \rightarrow \Psi_{\mu} H} = -\frac{\alpha_2^2 m_{\tau}^2 s t}{6 m_{3/2}^2 M_P^2 v^2}\,,
\end{equation}

\begin{equation}
     |\mathcal{M}|^2_{H H \rightarrow \Psi_{\mu} \ell} = -\frac{\alpha_2^2 m_{\tau}^2}{6 m_{3/2}^2 M_P^2 v^2} t(s +t)\,,
\end{equation}

\begin{equation}
    |\mathcal{M}|^2_{H \bar{H} \rightarrow \bar{\Psi}_{\mu} \Psi_{\mu}} = -\frac{\alpha_2^4}{96}\frac{t(s + t)^3}{m_{3/2}^4 M_P^4}\,,
\end{equation}

\begin{equation}
    |\mathcal{M}|^2_{L \bar{L} \rightarrow \bar{\Psi}_{\mu} \Psi_{\mu}} = \frac{\alpha_2^4}{128}\frac{t^4}{m_{3/2}^4 M_P^4}\,.
\end{equation}

\section{Loop Calculations}
\label{app:loops}
First, we consider the inflaton decay to two Higgs bosons through the loop process shown in~Fig.~\ref{brl1}. The amplitude is given by
\begin{align}
\notag
    \mathcal{M}_{\Phi \rightarrow HH} = & \, A_{\Phi H H} \int \frac{d^4 q}{(2 \pi)^4}
    \\
    & \times \frac{P_L \slashed{q} ( \slashed{q}+\slashed{p}_1 + M_R) ( \slashed{q}+\slashed{p}_2 + M_R)}{D_0 D_1 D_2},
    \label{app:higgs}
\end{align}
with the coupling $A_{\Phi H H} = -2 y_{\nu} y^2$, and the propagators are defined as $D_0 = q^2 - m_1^2 \simeq q^2$ and $D_i = (q + p_i)^2 - M_R^2 \, (i = 1, 2)$, where $m_1$ is the left-handed neutrino mass and $M_R$ is the right-handed neutrino mass. We remind the reader that when $m_{\Phi} > M_R$, we use the coupling $y_{\nu} = y_{\Phi}$. 

To calculate the amplitudes, we use the Passarino-Veltman functions~\cite{Passarino}. The two-point form factors can be expressed as
\begin{equation}
    B_0; B_{\mu}; B_{\mu \nu} = \int \frac{d^{4} q}{i \pi^{2}} \frac{1 ; q_{\mu} ; q_{\mu} q_{\nu}}{D_{0} D_{1}},
\end{equation}
where
\begin{equation}
    B_{\mu} = p_{1 \mu} B_1
\end{equation}
and
\begin{equation}
    B_{\mu \nu} = g_{\mu \nu} B_{00} + p_{1 \mu} p_{1 \nu} B_{11},
\end{equation}
and the three-point form factors are given by
\begin{equation}
C_0; C_{\mu}; C_{\mu \nu}; C_{\mu \nu \alpha}=\int \frac{d^{4} q}{i \pi^{2}} \frac{1 ; q_{\mu} ; q_{\mu} q_{\nu} ; q_{\mu} q_{\nu} q_{\alpha}}{D_{0} D_{1} D_{2}},
\end{equation}
where 
\begin{equation}
C_{\mu}=p_{1 \mu} C_{1}+p_{2 \mu} C_{2},
\end{equation}
\begin{align} \notag
C_{\mu \nu}=& g_{\mu \nu} C_{00}+p_{1 \mu} p_{1 \nu} C_{11}+p_{2 \mu} p_{2 \nu} C_{22} \\
&+\left\{p_{1 \mu} p_{2 \nu}+p_{2 \mu} p_{1 \nu}\right\} C_{12},
\end{align}
and
\begin{align} \notag
& C_{\mu \nu \alpha}= \sum_{i=1,2}\left\{g_{\mu \nu} p_{i \alpha}+g_{\nu \alpha} p_{i \mu}+g_{\alpha \mu} p_{i \nu}\right\} C_{00 i} \\ \notag
& + p_{1 \mu} p_{1 \nu} p_{1 \alpha}C_{111} +  p_{2 \mu} p_{2 \nu} p_{2 \alpha}C_{222} \\ \notag
&+\left\{p_{1 \mu} p_{1 \nu} p_{2 \alpha}+p_{1 \mu} p_{2 \nu} p_{1 \alpha}+p_{2 \mu} p_{1 \nu} p_{1 \alpha}\right\} C_{112} \\
&+\left\{p_{2 \mu} p_{2 \nu} p_{1 \alpha}+p_{2 \mu} p_{1 \nu} p_{2 \alpha}+p_{1 \mu} p_{2 \nu} p_{2 \alpha}\right\} C_{122}.
\end{align}
Using the Passarino-Veltman functions, we can express the amplitude~(\ref{app:higgs}) as
\begin{align} \notag
& \mathcal{M}_{\Phi \rightarrow HH}  = -\frac{i A_{\Phi H H}}{16 \pi^2} P_L [(-\slashed{p}_1 + \slashed{p}_2 +2 M_R) B_0 + \slashed{p}_2 B_1 \\ \notag &
+(M_R p_1^2 + p_1^2 \slashed{p}_2 + p_2^2 \slashed{p}_1 + 2 M_R^2 \slashed{p}_1 + M_R \slashed{p}_1 \slashed{p}_2)C_1 \\
& + (M_R p_2^2 - p_2^2 \slashed{p}_1 + 2 p_1 \cdot p_2 - p_1^2 \slashed{p}_2 - M_R \slashed{p}_1 \slashed{p}_2)C_2
]\,.
\end{align}
Assuming $M_R \gg m_{\Phi} \gg m_{H}$  
we obtain
\begin{equation}
    |\mathcal{M}_{\Phi \rightarrow HH}|^2 = \frac{y_{\nu}^2 y^4 M_R^2}{8 \pi^4} \ln^2 \left(\frac{M_R^2}{m_{\Phi}^2} \right),
\end{equation}
and the decay rate is given by Eq.~(\ref{phitoH}).

Next, we calculate the inflaton decay rate to left-handed neutrinos through the loop process shown in Fig.~\ref{brl1}. The amplitude of this process is
\begin{equation}
\begin{aligned}
\mathcal{M}_{\Phi \rightarrow \nu_L \nu_L}  = & \, A_{\Phi \nu_L \nu_L} \int \frac{d^4 q}{(2 \pi)^4} \bar{u}(p_1) P_R \\
    & \frac{( \slashed{q}+\slashed{p}_1 + M_R) ( \slashed{q}+\slashed{p}_2 + M_R)}{D_0 D_1 D_2} P_L v(p_2),
    \label{app:neut}
\end{aligned}
\end{equation}
where $A_{\Phi \nu_L \nu_L} = -2 y_{\nu} y^2$, $D_0 = q^2$, and $D_i = (q + p_i)^2 - M_R^2 \, (i = 1, 2)$.
Using the Passarino-Veltman functions, we can express the amplitude~(\ref{app:neut}) as
\begin{align} \notag
\mathcal{M}_{\Phi \rightarrow \nu_L \nu_L}  = &-\frac{i A_{\Phi \nu_L \nu_L}}{16 \pi^2} \bar{u}(p_1) P_R \\ \notag
&\times [
(M_R(\slashed{p}_1 + \slashed{p}_2 + M_R) + \slashed{p}_1 \slashed{p}_2)C_0 \\ \notag
& + (2 M_R \slashed{p}_1 + p_1^2 + \slashed{p}_1 \slashed{p}_2) C_1 \\ 
& + (2 M_R \slashed{p}_2 + p_2^2 + \slashed{p}_1 \slashed{p}_2 ) + B_0
] P_L v(p_2).
\end{align}
With $p_1^2 = p_2^2 = m_1^2$, $p_1 \cdot p_2 = \frac{m_{\Phi}^2}{2} - m_1^2$, and $M_R \gg m_{\Phi}$, we find 
\begin{equation}
    |\mathcal{M}_{\Phi \rightarrow \nu_L \nu_L}|^2 = \frac{y_{\nu}^2 y^4}{128 \pi^4} \frac{m_1^2 M_{\phi}^2}{M_R^2},
\end{equation}
and upon substitution of Eq.~(\ref{Eq:m1}), the decay rate is given by Eq.~(\ref{phitonuL}).

Finally, we calculate the inflaton decay rate to raritrons through the loop process shown in Fig.~\ref{brl2}. We can express the amplitude as follows,
\begin{align} \notag
   &\mathcal{M}_{\Phi \rightarrow \Psi_{\mu} \Psi_{\mu}} =  -\frac{i A_{\Phi \Psi_{\mu} \Psi_{\mu}}}{16 \pi^2} \int \frac{d^4 q}{(2 \pi)^4} \bar{u}_{\mu}(p_1) [\gamma_{\rho}, \slashed{q}] \gamma^{\mu} \\
    & \times \frac{(\slashed{q}+\slashed{p}_1 + M_R) ( \slashed{q}+\slashed{p}_2 + M_R)}{D_0 D_1 D_2} \gamma^{\nu}  [\slashed{q}, \gamma^{\rho}] v_{\nu} (p_2) ,
\label{app:rar}
\end{align}
where $A_{\Phi \Psi_{\mu} \Psi_{\mu}} = \frac{2 \alpha_1^2 y_{\nu}}{M_P^2}$, $D_0 = q^2$, and $D_i = (q + p_i)^2 - M_R^2~(i = 1,2)$. We do not include the full expression of the amplitude~(\ref{app:rar}) in terms of the Passarino-Veltman functions due to its complexity. 
With $M_R \gg m_{\Phi}\gg m_{3/2}$, the amplitude takes the form
\begin{equation}
    |\mathcal{M}|^2 = \frac{2 \alpha_1^4 y_{\nu}^2 m_{\Phi}^2}{9 \pi^4 M_p^4 m_{3/2}^4} \left[5 - 6 \ln \left(\frac{M_R^2}{m_{\Phi}^2}\right) \right]^2,
\end{equation}
and the decay rate is given by Eq.~(\ref{phitorar}).


\vspace{-.5cm}
\bibliographystyle{apsrev4-1}

\end{document}